\documentclass{aa}  
\bibpunct{(}{)}{;}{a}{}{,} 
\usepackage{color}
\usepackage{graphicx}
\usepackage{txfonts}
\begin{document}

   \title{Low temperature MIR to submillimeter mass absorption coefficient of interstellar dust analogues I:  Mg-rich glassy silicates}


   \author{
   	  K. Demyk\inst{1}
 \and  C. Meny\inst{1}
 \and  X.-H. Lu\inst{1}
 \and  G. Papatheodorou\inst{2,3}
 \and  M.J. Toplis\inst{1}
 \and  H. Leroux\inst{4}
 \and  C. Depecker\inst{4}
 \and  J.-B. Brubach\inst{5}
 \and  P. Roy\inst{5}
 \and  C. Nayral\inst{6}
 \and   W.-S. Ojo\inst{6}
 \and  F. Delpech\inst{6}
 \and  D. Paradis\inst{1}
 \and  V. Gromov\inst{7}
   }
  
 \institute{
IRAP, Universit\'e de Toulouse, CNRS, UPS ; IRAP; 9 Av. colonel Roche, BP 44346, F-31028 Toulouse cedex 4, France
\and
Institute of Chemical Engineering Sciences, FORTH, P.O. Box 1414, GR-26504 Patras, Greece
\and
Department of Chemical Engineering, University of Patras, GR-26504 Patras, Greece
\and
UMET, UMR 8207, Universit\'e Lille 1, CNRS, F-59655 Villeneuve d'Ascq, France
\and
Ligne AILES - Synchrotron SOLEIL, L\'\ Orme des Merisiers, F-91192 Gif-sur-Yvette, France 
\and
Universit\'e de Toulouse, INSA, CNRS, LPCNO, F-31077 Toulouse, France
\and
Space Research Institute, RAS, 84/32 Profsoyuznaya, 117810 Moscow, Russia  
 }

   \date{Received ...; accepted ...}

 
  \abstract
   {The submillimeter spectral domain has been extensively explored by the {\it Herschel} and {\it Planck} satellites and is now reachable from the ground with ALMA. A wealth of data, revealing cold dust thermal emission, is available for astronomical environments ranging from interstellar clouds, cold clumps, circumstellar envelops, and protoplanetary disks. The interpretation of these observations relies on the understanding and modeling of cold dust emission and on the knowledge of the dust optical properties. }
   {The aim of this work is to provide astronomers with a set of spectroscopic data of realistic interstellar dust analogues that can be used to interpret the observations. It pursues the experimental effort aimed at characterizing the spectroscopic properties of interstellar dust analogues at low temperature in the mid infrared (MIR) to millimeter spectral domain. Compared to previous studies, it extends the range of studied dust analogues in terms of composition and of structure of the material. }
   {Glassy silicates of mean composition (1-$x$)MgO - $x$SiO$_2$ with $x$ = 0.35 (close to forsterite, Mg$_2$SiO$_4$), 0.50 (close to enstatite, MgSiO$_3$) and 0.40 (close to Mg$_{1.5}$SiO$_{3.5}$ or MgSiO$_3$:Mg$_2$SiO$_4$ = 50:50) were synthesized. The mass absorption coefficient (MAC) of the samples was measured in the spectral domain 30 - 1000 $\mu$m for grain temperature in the range 300 K - 10 K and at room temperature in the 5 - 40 $\mu$m domain. }
   {We find that the MAC of all samples varies with the grains temperature and that its spectral shape cannot be approximated by a single power law in ${\lambda}^{-\beta}$. In the FIR/submm, and above 30K, the MAC value at a given wavelength increases with the temperature as thermally activated absorption processes appear. The studied materials exhibit different and complex behaviors at long wavelengths ($\lambda \geq$ 200 to 700 $\mu$m depending on the samples).These behaviors are attributed to the amorphous nature of dust and to the amount and nature of the defects within this amorphous structure. We do not observe MAC variations in the 10-30 K range. Above 20 $\mu$m, the measured MAC are much higher than the MAC calculated from interstellar silicate dust models indicating that the analogues measured in this study are more emissive than the silicates in cosmic dust models. }
{The underestimated value of the MAC deduced from cosmic dust models in the FIR/submm has important astrophysical implications because masses are overestimated by the models. Moreover, constraints on elemental abundance of heavy elements in cosmic dust models are relaxed. }


\titlerunning{Low temperature MIR/submm mass absorption coefficient of Mg-rich glassy silicates}
\authorrunning{K. Demyk et al.}
\maketitle

%

\section{Introduction}
 \label {intro}
 
 In astrophysical observations, the far-infrared (FIR) and submillimeter (submm) emission ($\sim$ 100 $\mu$m -  2mm) is shown to be dominated by a component from big nanograins (BG) with radius in the range $\sim$ 10-15 nm  to $\sim$ 100 - 400 nm, big enough to radiate at thermal equilibrium with the surrounding radiation field. Assuming optically thin emission which is often a valid hypothesis in the FIR-submm domain, the dust spectral energy distribution (SED) observed through a few broadband filters appears to be coherent with the following modified blackbody (MBB) emission law :

\begin{equation}
\label{modified_bb}
\mathrm{ I_{\lambda} = {\tau}_{\lambda} \: B_{\lambda}(T_{dust}) = M_{dust}  \: d^{-2}  \:  {\Omega}^{-1} \: {\kappa}_{\lambda} \: B_{\lambda}(T_{dust})}
\end{equation}

\noindent where ${\tau}_{\lambda}$ is the dust optical depth, and $\mathrm{B_{\lambda}(T_{dust})}$ is the Planck blackbody emission law at the dust color temperature $\mathrm{T_{dust}}$, M$_{d}$ is the dust mass along the line of sight, d the distance of the source, $\Omega$ the beam solid angle and  ${\kappa}_{\lambda}$ the mass absorption coefficient of the dust along the line of sight (MAC, cm$^2$.g$^{-1}$). ${\kappa}_{\lambda}$ is a fundamental quantity which contains all the information on dust: its composition and structure (amorphous or crystalline, porosity, coagulation), its size distribution and the way it interacts with light.  Simple semi-classical physical models for bulk material provide, in the FIR/submm domain, a temperature independent asymptotic behavior for the dust emissivity which is widely used in astronomical modeling and is expressed as follows:  

\begin{equation} \label{kappa}
\mathrm{\kappa_{\lambda} =\:\kappa _{\lambda_{0} }\: ( \frac{\lambda}{\lambda}_{0}) ^{-\beta}}
\end{equation}  

\noindent where the spectral index $\beta$ has a constant value (i.e. temperature and wavelength independent value) that ranges from one to two. The value  $\beta$=2 is commonly adopted because it is the asymptotic value toward long wavelengths of the two most simple models for dust absorption in the FIR-submm : the Lorentz oscillator model for absorption in crystals and the Debye model (with a constant optical coupling coefficient) for absorption in 3 dimensional amorphous structures \citep{henning1997}. 

Historically, the first observations by the ballon-borne missions {\it PRONOAS} and {\it Archeops} in the FIR/submm wavelength range, revealed an unexpected result : an anticorrelation between the observed spectral index ${\beta}_{obs}$ and the temperature $\mathrm{T_{dust}}$  \citep{dupac2003,desert2008}. More recently, the observations carried out by the {\it Herschel} (160 - 550 $\mu$m) and {\it Planck} missions (160 $\mu$m - 3 mm for the HFI instrument) have greatly increased the number and precision of available data. Combined with the FIRAS 100 $\mu$m band, these observations allow astronomers to build spectral energy distribution (SED) from the FIR to the millimeter range with unprecedented sensitivity, spectral resolution for {\it Herschel} observations and over the whole sky with the {\it Planck} mission. In all these observations, the ${\beta}_{obs}$ - $\mathrm{T_{obs}}$ anti-correlation is observed \citep{planck2011_early_XXV,planck2011_early_XXIII,planck2013-XI,planck-XVII-Int-2014,malinen2014,juvela2015}. Depending on the environments, this $\beta$ - T anti-correlation is observed on a temperature range of 10 to 30 K, up to 60 K if hotter clouds are considered. Meanwhile, many studies have been conducted to determine all possible causes for the variation of ${\beta}_{obs}$ with $\mathrm{T_{obs}}$: the effect of noise in the data and of the choice of the method used to fit the observations \citep{shetty2009b,juvela2012a,juvela2013}; the effect of specific distributions of density and temperature profiles of interstellar grains along the line of sight \citep{malinen2011,juvela2012b,pagani2015}; the intrinsic properties of grains \citep{meny2007,koehler2012,paradis2011,jones2012a,jones2012b,jones2013}.

However, up to now the FIR and submm observations reveal some facts that remain valid, when interpreted with the MMB emission law :

i) The dust spectral index and temperature are statistically anticorrelated, as discussed previously in this introduction.

ii) The spectral index value ${\beta}_{obs}$  and the mass absorption coefficient ${\kappa}_{{\lambda}_0}$ evolve in the different phases of the ISM (from diffuse to dense media, from HI regions to cold molecular clouds, with metallicity). At the Galactic scale, \cite{paradis2012} observe in {\it Herschel} data that the submm excess (or lowering of the spectral index) increases with the Galactic longitude. In extragalactic environments, the Large and especially the Small Magellanic clouds (respectively LMC and SMC) have a pronounced FIR and submm excess compared to the Milky Way \citep{galliano2011,bot2010}. It is also observed that the FIR-submm emissivity, normalized to the visible or the near-infrared extinction, increases by a factor greater than 3 and that the temperature decreases by 10-20\%, from diffuse medium to denser ISM \citep{stepnik2003, koehler2012}. More recently, \cite{paradis2014}, using 70 - 500 $\mu$m Herschel data combined with 1.1 mm Bolocam data, show that the FIR-submm emission in ultra-compact HII regions and in cold molecular clumps are characterized by two different spectral index values.

iii) The dust spectral index varies with wavelength. This observational fact has been often recognized and is referred to as submm flattening of the SED or submm excess. It has been at the origin of two interpretations : {\it (i)} The Finkbeiner two big grain components model, in which each component gives rise to a FIR emission following equations (1) and (2), the coldest component achieving the submm excess \citep{Finkbeiner1999, Meisner2015}. {\it (ii)} The TLS model in which the variation of the spectral index (with wavelength and temperature) is linked to intrinsic properties of the amorphous state of only one interstellar dust component \citep{meny2007,paradis2011}. The TLS model successfully reproduces the dust emission in ultra-compact HII regions and cold molecular clumps with the same dust component in both types of environment without the need to consider two different dust components associated with each environment (and characterized by different spectral index values) as is the case in the MBB emission model \citep{paradis2014}. An analysis of the outer Galactic plane using DIRBE, Archeops and WMAP data reveals as a general phenomenon a change in the emissivity around $\lambda$ = 600 $\mu$m in the SED, with an emission steeper in the FIR than in the submm and mm, both in molecular clouds and atomic phases \citep{paradis2009}. Such wavelength dependence of the spectral index is also well demonstrated by the study of \cite{juvela2015} who determine reliable estimates of the mean spectral index (using the MBB model) in 116 cold cores selected from the {\it Planck} Cold Clump Catalogue C3PO and mapped with {\it Herschel} PACS and SPIRE instruments. Using different combinations of photometric data, \citet{juvela2015} showed that the best fit to the observations are obtained with two components. In the absence of knowledge of the wavelength dependence of the spectral index, they adopted a wavelength threshold value of $\lambda$ = 700 $\mu$m, and determined  a millimetric spectral index ${\beta}_{obs}^{mm}$ = 1.66 close to the value observed by Planck in the diffuse medium \citep[${\beta}_{obs}^{mm}$ = 1.51,][]{planck-XVII-Int-2014}, and a FIR spectral index ${\beta}_{obs}^{FIR}$ = 1.91. Using only Herschel data the authors found that many individual cold clumps give values up to ${\beta}_{obs}^{FIR} \sim 2.2$, and pointed out that temperature gradients in optically thick sources can only decrease the FIR spectral index value and that in consequence the FIR spectral index value may even be steeper in the limited wavelength range covered by Herschel.

From the modeling point of view, dust models aim to reproduce astrophysical observables (extinction, emission, MIR bands, polarization, etc.) over the widest spectral range (from far-ultraviolet, FUV, to the millimeter range). They should be able to propose a coherent scenario of the evolution of dust properties (composition, structure, and coagulation, amongst others) through various environments (diffuse ISM, molecular clouds, PDRs and HII regions, different metallicities, etc.) and be consistent with the expected ISM abundance of the elements in the various environments. All dust models \citep[see for example][]{compiegne2011,jones2013,draine2007} consider at least two dust components: a component of very small grains (VSG) with grain radius a $\leq$ 10-15 nm, and a big grain (BG) component. The VSG component is dominated by carbon grains (either graphite or amorphous carbon more or less hydrogenated, denominated hereafter as a-C(:H)), transiently heated by the ISRF and dominating the near and mid-infrared with a residual tail in the FIR-submm. The BG component drives the FIR-submm emission. Although silicates dominate the BG component in terms of mass, some mixing with the carbonaceous phase is expected, and thus the FIR-submm SEDs reflect optical properties of both the silicate and carbonaceous phases. According to recent dust models such mixing could occur, either by introducing a complementary component of pure a-C( :H) big grains, or by accretion of mantles or aggregation processes inside or between the a-Sil and a-C(:H) phases \citep{jones2013}. In that case, ${\kappa}_{\lambda}$ in Eq.~\ref{modified_bb} is thus averaged over the two BG components. 

The interpretation of the well established observational facts in the FIR-submm described above, requires detailed knowledge of the various expected silicate and carbonaceous components of the ISM. Furthermore, the robustness of dust models relies on accurate knowledge of the optical properties of those components over the whole FUV-mm wavelength range. These optical properties have to be determined from relevant combinations of laboratory measurements performed on well characterized (e.g. stoichiometry, atomic-structure and grain sizes) interstellar dust analogues (silicates, carbonaceous compounds, ices) or pure elements or compounds (graphite, Fe, Mg, FeS, MgO,…), from physical models of interaction between the solid-state and electromagnetic waves, and from astrophysical observational constrains.

Within this context, this work presents a new experimental study of silicate dust analogues. We have measured the MAC of magnesium rich glassy silicate grains with composition of enstatite (MgSiO$_3$), of composition close to forsterite (Mg$_2$SiO$_4$) and a non-stoichiometric composition (MgSiO$_3$:Mg$_2$SiO$_4$ = 50:50) in the 5 $\mu$m - 1 mm spectral domain and for grain temperature varying from room temperature to 10K. This study extends our work on magnesium rich amorphous silicates synthesized by solgel methods, but for which spectroscopic measurements were limited to wavelengths longer than 100 $\mu$m \citep{coupeaud2011}. \citet{boudet2005} also measured the MAC of glassy and solgel silicates in the wavelength range 100 $\mu$m  - 1 mm while \cite{mennella1998} measured iron rich amorphous silicates in the range 20  $\mu$m - 2 mm. All these studies show that the MAC of dust grains varies with temperature: at a given wavelength the MAC decreases as T decreases. This decrease of the MAC is more pronounced at long wavelengths and consequently the wavelength dependence of the MAC can no longer be described by an asymptotic power law with a single spectral index, as assumed in dust models.

This paper is organized as follows. Sect.~\ref{studied_samples} describes the studied dust analogues (synthesis and characterization), the experimental setups used for the measurements, the sample preparation and the data reduction performed to reconstruct the MAC. The MAC measurements of each samples are presented in Sect.~\ref{results}. They are compared with previous experimental data in Sect.~\ref{comp_lab} and with astronomical dust models in Sect.~\ref{discuss} which also discuss the astrophysical implications of these new measurements.

\section{Experiments}

\subsection{Samples synthesis and characterization}
\label{studied_samples}

The studied samples are Mg-rich silicate glasses synthesized by two different methods. The first method is described in details in \citet{kalampounias2009}. Briefly, it consists in putting a sintered mixture of the oxides having the desired stoichiometry  in a controlled Ar gas flow jet and using a CO$_2$ laser for heating and melting. The heating of the levitated sample must be controlled to avoid overheating. The sample is melted for less than 5 s and when the laser is shut down amorphous spherical samples of $\sim$ 1 - 1.5 mm diameter, colorless and transparent, are produced. In the second method, appropriate proportions of silicon dioxide and magnesium carbonate powders is heated together in a platinum crucible in air at 1600$^ {\circ}$C for approximately 30 minutes. The liquid is quenched to glass by dipping the bottom of crucible in water. This glass was then crushed and remelted at the same conditions to ensure homogeneity. 
The four samples studied in this work have mean compositions of (1-$x$)MgO-$(x)$SiO$_2$ with $x$ = 0.35, 0.40, 0.50. The mean composition $x$ = 0.35 is close to forsterite (Mg$_2$SiO$_4$, i.e. $x$ = 33.3), while $x$ = 0.50 is close to a mean composition of enstatite (MgSiO$_3$) and $x$ = 0.40 to an intermediate composition in between MgSiO$_3$ and Mg$_2$SiO$_4$. Three samples were synthesized with the first method: X35, X40 and X50A, corresponding to $x$ = 0.35, 0.40 and 0.50 respectively and one sample synthesized with the second method: X50B, corresponding to $x$ = 0.50. The X50A and X50B samples have the same stoichiometry but, they were synthesized by two different methods.Both have been studied to verify that synthesis technique does not influence microscopic structure.  

The structural and chemical local order within the network in silicate glasses may be characterized by the amount of Q$^\mathrm{i}$ species, with i=0,1,2,3,4 designating the number of bridging oxygens around a given silicon atom. The Q$^\mathrm{i}$ value can be derived from Raman or $^{29}$Si and $^{25}$Mg MAS NMR spectroscopic studies. Q$^\mathrm{0}$ species correspond to isolated tetrahedra, SiO$^{4-}_4$, Q$^\mathrm{1}$ species to dimers Si$_2$O$^{6-}_7$, and Q$^\mathrm{2}$ species to chains. A glass made of Q$^\mathrm{3}$ and Q$^\mathrm{4}$ species is highly interconnected and may have diverse 3-dimensional structures \citep[see Fig. 5 in][]{kalampounias2009}.  \citet{sen2009} have determined the relative proportion of Q$^\mathrm{i}$ species in 9 glasses from a series of (MgO)$_y$(SiO$_2$)$_{100-y}$ glasses with $y$ in the range 50-66.7 ($y$=50, 60, 66.7 correspond to $x$=0.5, 0.40, 0.33 in the notation adopted here). The samples X35, X40 and X50A were synthesized by the same team using the same technique as the study of Sen et al. (2009) allowing direct comparison of the results presented here with those of Sen et al. (2009). From a structural point of view the studied samples mainly differ by their degree of polymerization, i.e, their amount of shared SiO$^{4-}_4$ tetrahedra. \citet{sen2009} find that the X50A sample contains Q$^\mathrm{1}$, Q$^\mathrm{2}$, Q$^\mathrm{3}$ and Q$^\mathrm{4}$ species in respective proportions of 25.0, 42.0, 25.7 and 7.3\%, whereas the X40 samples contains Q$^\mathrm{0}$, Q$^\mathrm{1}$ and Q$^\mathrm{2}$ species in respective proportions of 27.5, 58.0 and 14.5\% while the X35 sample contains only Q$^\mathrm{0}$ and Q$^\mathrm{1}$ species in respective proportions of about 50-60 and 50-40\%. As expected, the X35 sample close to the forsterite composition, is less polymerized than the the X50A sample, of enstatite composition, the X40  being intermediate.

\subsection{Spectroscopic measurements}
\label{spectro}

The MAC of the samples were recorded on the ESPOIRS setup at IRAP (Institut de Recherche en Astrophysique et Planétologie) and on the AILES (Advanced Infrared Line Exploited for Spectroscopy) beam line at the synchrotron SOLEIL \citep{brubach2010}. The ESPOIRS setup consists in a Fourier Transform InfraRed (FTIR) spectrometer (Bruker Vertex 70V) coupled with a closed cycle optical cryostat (QMC Instruments) allowing to cool down the samples from 300 K to 10 K. Thanks to the combination of various sources (Globar, Hg and tungsten lamps), beam splitters (CaF$_2$, CsI, Si) and detectors (InGasAs, DLaTGS, closed cycle He cooled transition edge superconducting bolometer from QMC Instrument)  the spectrometer covers the spectral domain from 0.7 $\mu$m to 1 mm. The sample cryostat contains a rotating wheel on which six samples may be mounted and measured without having to break the vacuum. At SOLEIL, the experimental setup of the AILES beam line is composed of a Bruker FTIR 125V spectrometer equipped with a Mylar 125 $\mu$m beamsplitter, a 1.6K Si bolometer and a closed-cycle sample cryostat (300K - 10K). The synchrotron radiation is much brighter than internal light sources and thus allows to obtain more reliable spectra in the 700 - 1000 $\mu$m range than internal sources. Measurements at wavelengths greater than 400 $\mu$m up to 1mm and more were performed at SOLEIL which allowed a good overlap of the two sets of measurements. For $\lambda \le 40 \mu$m the measurements were performed at room temperature and for $\lambda \ge 30 \mu$m they were performed at 300 K, 200 K, 100 K, 30 K, and 10 K.\\

The X35, X40 and X50A samples obtained by levitation and  CO$_2$ laser melting are spheres of diameter of $\sim$ 1mm, the X50B sample is bulk transparent material. The four samples were ground by hand in an agate mortar until (sub)micronic sizes are reached. For the transmission measurements, the samples are prepared in the form of pellets. For the MIR domain (2 - 40 $\mu$m), $\sim$ 0.5-1.0 mg of sample were mixed with 300 mg of KBr or CsI (Aldrich) for several minutes in an agate mortar and pressed under ten tons for five minutes to produce a 1.3 mm diameter pellet. For measurements above 40 $\mu$m, KBr and CsI are not transparent and are replaced by polyethylene (PE, Thermo Fisher Scientific). The PE and sample mixture were heated on a heating plate up to 130$^ {\circ}$C (measured on top of the dye) for five minutes before pressing under ten tons for five minutes to produce a 1.3 mm diameter pellet. Because the MAC of the samples decreases at long wavelengths, several pellets with an increasing mass of samples were used. Typically, pellets containing 5, 20, 40 80 and 170 mg of samples and 300 mg of PE were retained.


\subsection{Data reduction}
\label{data_red}
 
The MAC of the pellet (sample and matrix), $\mathrm{{\kappa}_{pellet}}$ (cm$^2$.g$^{-1}$), is related to the transmittance spectrum as follows \citep[e.g.][]{bohren1998}: 
\begin{equation}\label{kappa-t}
\mathrm{{\kappa}_{pellet} = - \frac{S}{m} \times ln(T)}
\end{equation}
where S is the pellet area (cm$^2$), m the sample mass in the pellet (g) and T the transmittance spectrum. The MAC of the grains, $\kappa$, is deduced from the MAC of the pellet by the following relation \citep[Eq. (6) from ][]{mennella1998}: 

\begin{equation}\label{kappa-g}
\mathrm{{\kappa}(\lambda, f) = g(\epsilon, f) * {\kappa}_{pellet}(\lambda, f)}
\end{equation}

where f is the volume filling factor of the grains within the matrix (f never exceeds 0.15 for all the pellets) and g($\epsilon$, f) is the finite concentration reduction factor. g($\epsilon$, f) is calculated using the formula given by \cite{mennella1998} and using their value for the dielectric function of the host medium (PE) and for the grain material. This calculation revealed that g($\epsilon$, f) is always greater than 0.88. The transmittance corrected for reflexion loss, $\mathrm{T_{corr}}$, is calculated using the formula: $\mathrm{T_{corr} = T_{measured} /  (1-R)^2 }$ with R, the reflectance of the pellet, expressed as $\mathrm{R = (1-n_{eff})^2 / (1+n_{eff})^2}$. The effective refractive index of the pellet, $\mathrm{n_{eff}}$, is related to the average dielectric function of the pellet, $\mathrm{{\epsilon}_{eff}}$, by the formula $\mathrm{n_{eff} = \sqrt{Re({\epsilon}_{eff})}}$. The correction factor for reflexion loss, (1-R)$^2$, is larger than 0.9 for all the pellets. Such correction is based on the assumption that the grains are spherical. However this is not the case even though SEM images show that the grains have a spheroidal shape close to a sphere. To correct for the shape effect one has to make an assumption on the grain shape distribution and to model the resulting MAC spectrum from the optical constants of some material close to the studied one. The drawback of such an approach is that the correction factor will be dependent on the shape distribution and on the optical constants chosen and we have not applied it here. It has been estimated in \citet{mennella1998} to be 1.35 for the amorphous fayalite samples.

The overall MAC curve is built from the transmittances of pellets having different masses. Before extracting their respective response we suppress from the transmittance curves the fringes due to internal reflexions between the two internal faces of the pellet and to the fact that the blank and sample pellets do not have the same exact thickness. The procedure to suppress the fringes is similar to that used in \cite{coupeaud2011}. The resulting transmittances are then converted to MAC and the different MAC are joined together by averaging the spectra in the overlapping regions.

The uncertainty on the MAC has two origins. The first is the uncertainty on the transmittance due the spectrometer thermal stability ($ \mathrm { {\Delta}MAC_{stab} = ({\delta}T/T) * S / m}$). It is estimated to be $ \mathrm {{\delta}T/T }\sim$ 2.5\% above 40 cm$^{-1}$ and $\mathrm {{\delta}T/T }\sim$ 5\% below. The second is the uncertainty on the mass of sample contained in the pellet, $ \mathrm { {\Delta}MAC_m = MAC(\lambda) * {\delta}m/m }$. It becomes important in the MIR since the amount of sample is very low, typically less than 1 mg and the precision of the weighing scale is $\mathrm{ {\delta}m \simeq 0.1}$ mg. The error on the MAC is thus the quadratic sum of these two uncertainties. To estimate the total error on the measured kappa over the entire spectral range we have taken the highest uncertainty on the mass and on the spectrometer stability and applied it to the whole spectral range. 

The absolute value of the MAC was set in the spectral range where the error on the mass is minimized while the dilution factor of the sample in the pellet is high enough to prevent any significant inhomogeneity in the pellet. This corresponds to pellets containing about 2-5 mg of sample and a spectral region in the 40-50 $\mu$m range. 

\begin{figure}[!t]
\includegraphics[scale=.37, angle=90]{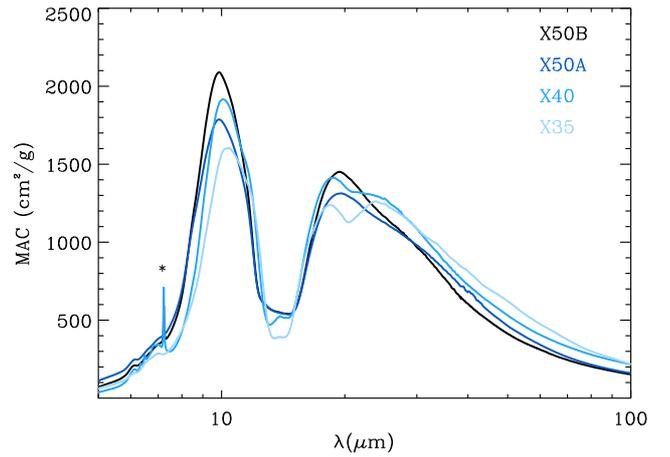}  
       \caption{Mass absorption coefficient of the samples X35, X40, X50A and X50B at room temperature in the MIR. The band labeled with the asterisk is not associated with the samples and is due to the PE matrix.   }
    \label{kappa_mir}
\end{figure}


\begin{figure*}[!t]
\includegraphics[scale=.4]{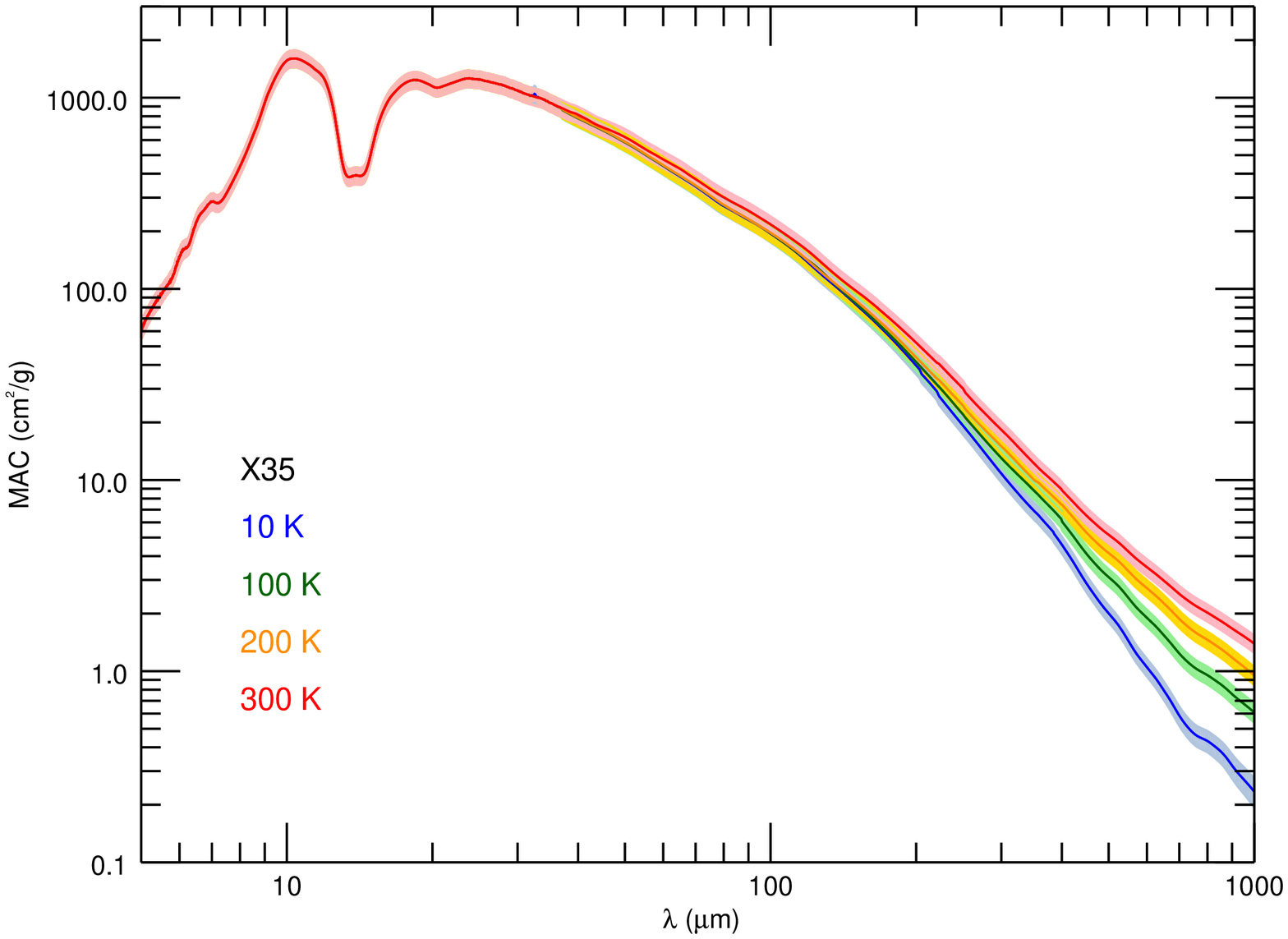}  
\includegraphics[scale=.4]{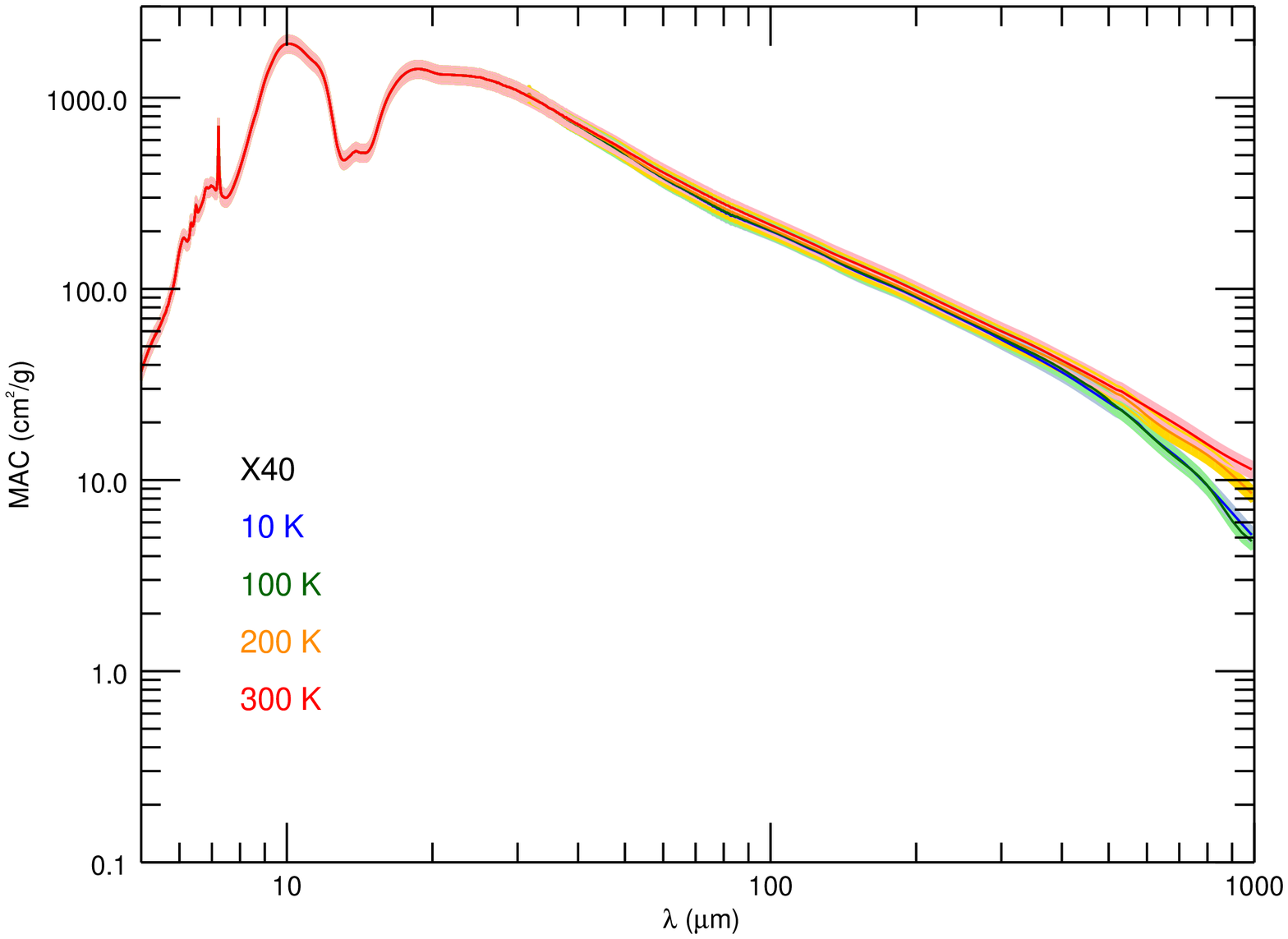}  
\includegraphics[scale=.4]{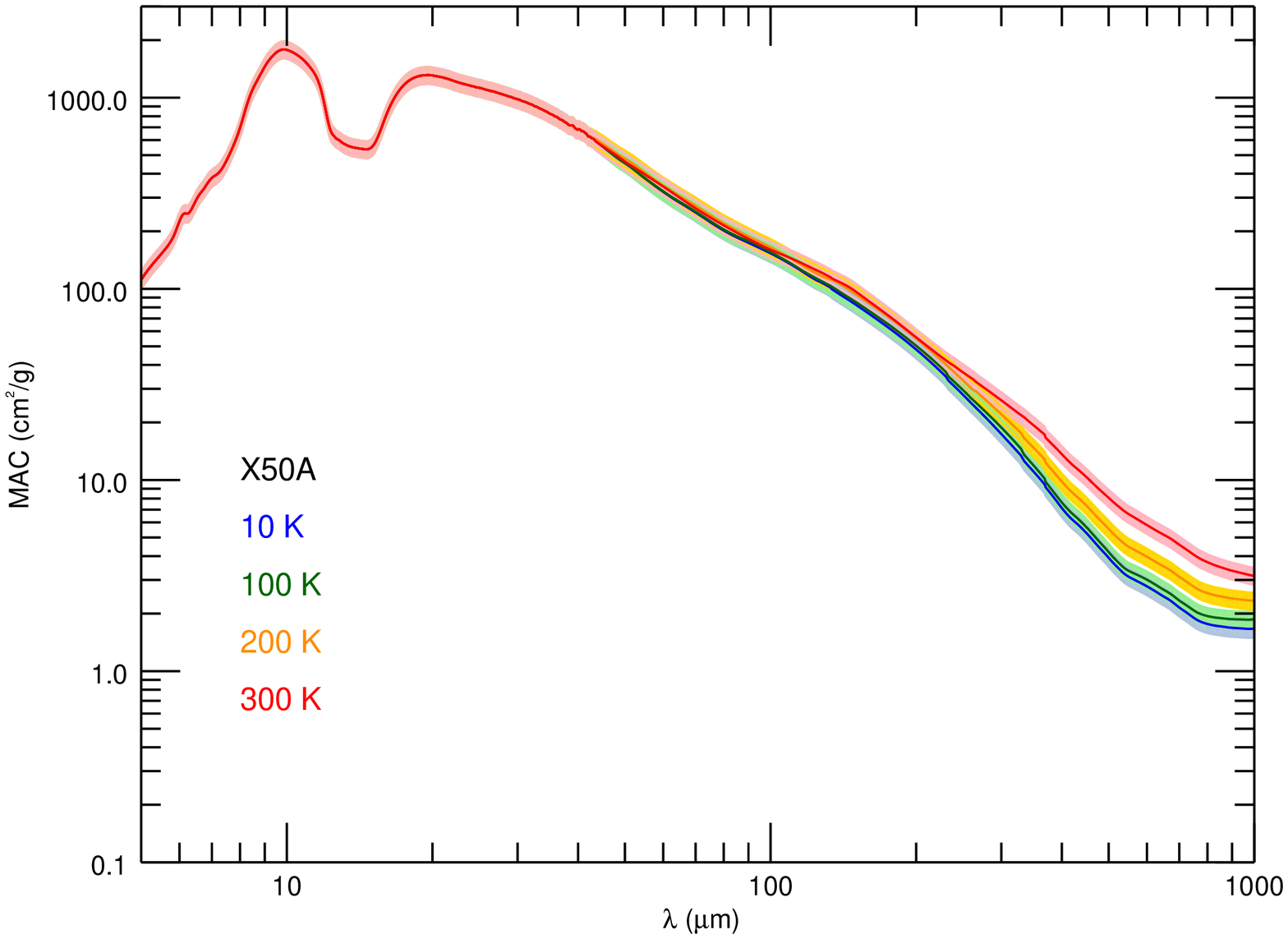}  
\includegraphics[scale=.4]{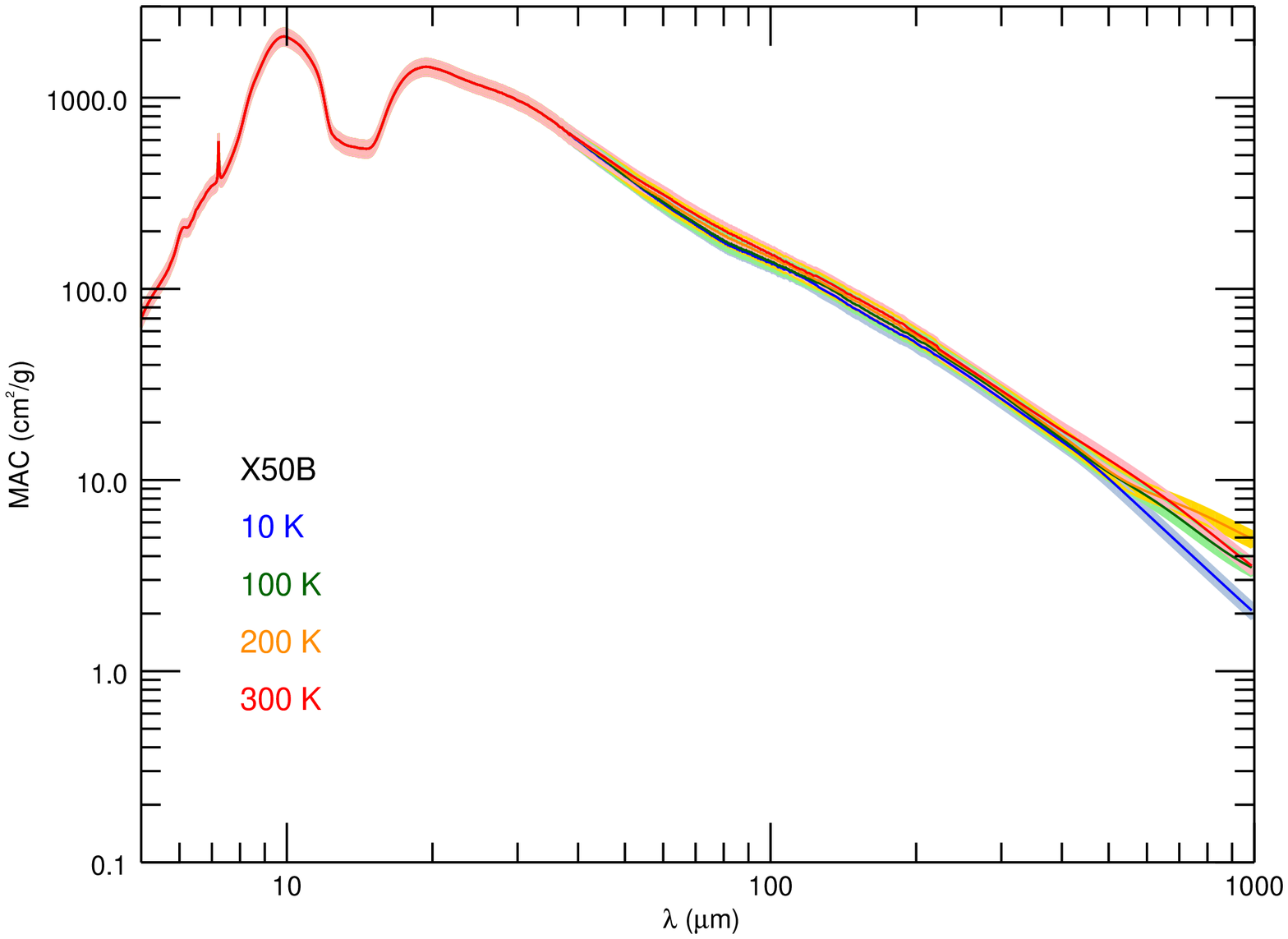}  
       \caption{Mass absorption coefficient of the samples X35 {\it (upper left panel)}, X40 {\it (upper right panel)}, X50A  {\it (lower left panel)} and X50B  {\it (lower right panel)}. For each sample we show the spectra measured for a grain temperature of 300 K (red), 200 K (orange), 100 K (green) and at 10 K (blue). The shaded area represents the uncertainty on the experimental data (see Sect.~\ref{spectro}). The sharp band at 7.2 $\mu$m is not associated with the samples and is due to the PE matrix.}
    \label{kappa_all}
\end{figure*}

\section{MIR and FIR/submm mass absorption coefficient as a function of the temperature}
\label{results}

In the MIR domain, the spectra show the two vibrational bands characteristic of amorphous silicates: the stretching mode of the Si-O bond at $\sim$ 10 $\mu$m and the bending mode of the O-Si-O bond at $\sim$ 19 $\mu$m (Fig.~\ref{kappa_mir}). The broadness and the absence of structure of the bands confirm the amorphous state of all four samples. The peak position of the stretching band shifts toward long wavelengths from the X50A and X50B samples (9.87 $\mu$m) to the X40 and X35 samples (10.10 $\mu$m and 10.34 $\mu$m, respectively). In addition, the stretching band of the X40 and X35 samples exhibits a shoulder at respectively 11.68 $\mu$m and 11.93 $\mu$m, that is not observed in the two X50 samples. The bending mode peaks at 19.42 $\mu$m for X50A and X50B samples whereas it shows two peaks for the X40 and X35 samples: the first one at 18.55 $\mu$m and the second one at 24.39 $\mu$m, the latter being more pronounced in the spectrum of the X35 sample than in the one of the X40 sample.The position and width of the vibrational bands depends on the polymerization of the SiO$_4$ tetrahedra within the sample which is related to the metal/Si ratio and to NBO/T, the average number of non bridging oxygen in each SiO$_4$ tetrahedra. The sample X35 containing the most MgO is the least polymerized sample whereas the samples X50 containing the least MgO are the most polymerized. Such a shift of the stretching mode toward long wavelengths with increasing content of MgO and the opposite behavior for the bending mode is in agreement with previous studies \citep{jager2003,mcmillan1984}. \citet{hofmeister1987} attributes a peak at 410 cm$^{-1}$ (24.39 $\mu$m) to rotation modes of the  SiO$_4$ tetrahedra in crystalline forsterite. This assignment is compatible with our results for glassy samples and the fact that this band is not observed in the X50 samples which contains no Q$^\mathrm{0}$ species (isolated tetrahedra, SiO$^{4-}_4$). The X50A and X50B samples have very similar MIR spectra as expected since they have the same stoichiometry but this also seems to show that they also have similar structural states. We note that the polymerization of the samples also has an impact on the FIR MAC as proposed by \citet{coupeaud2011} (see Sect.~\ref{discuss}).\\

\begin{figure}[!t]
\includegraphics[scale=.185, angle=90]{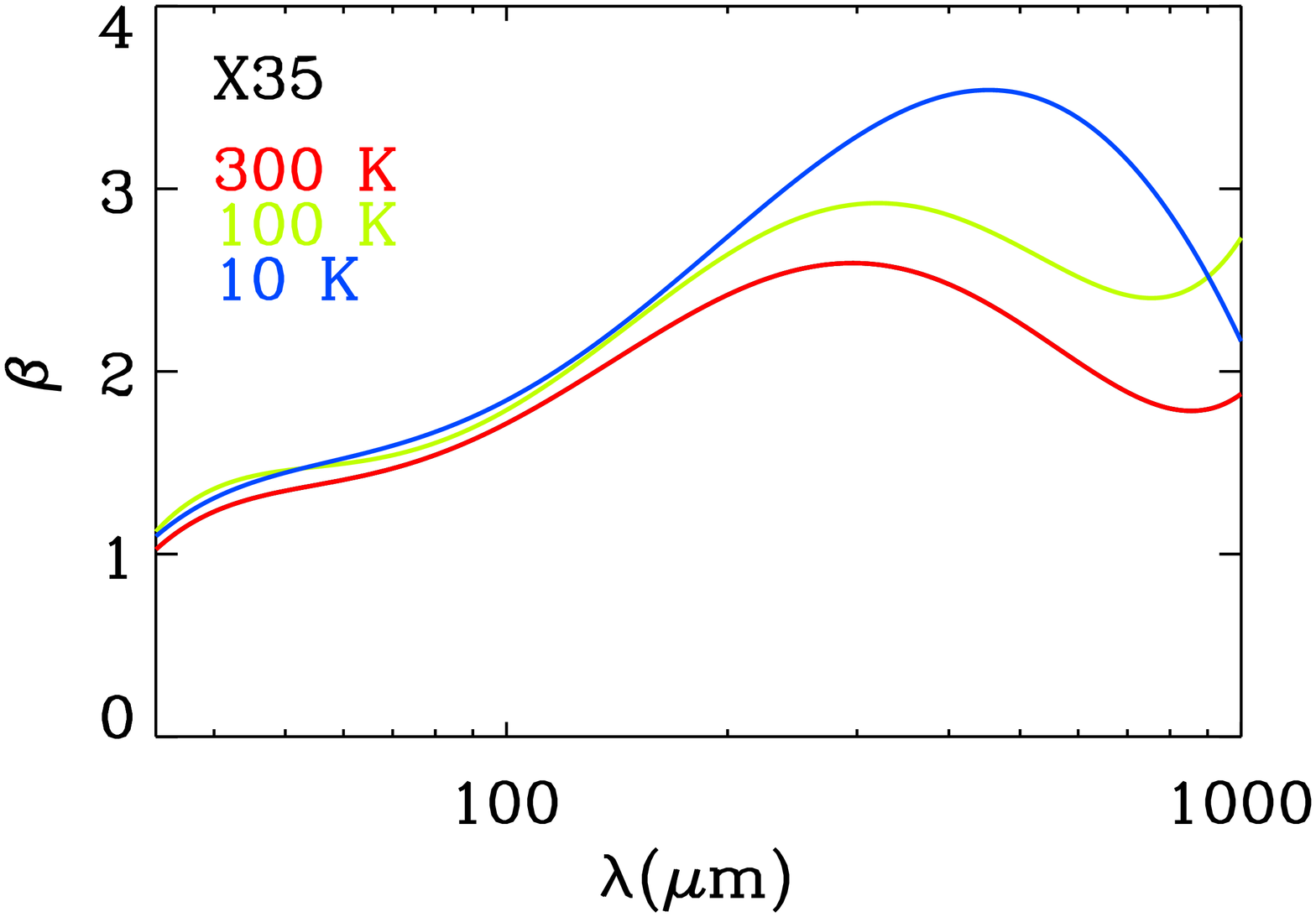}  
\includegraphics[scale=.185, angle=90]{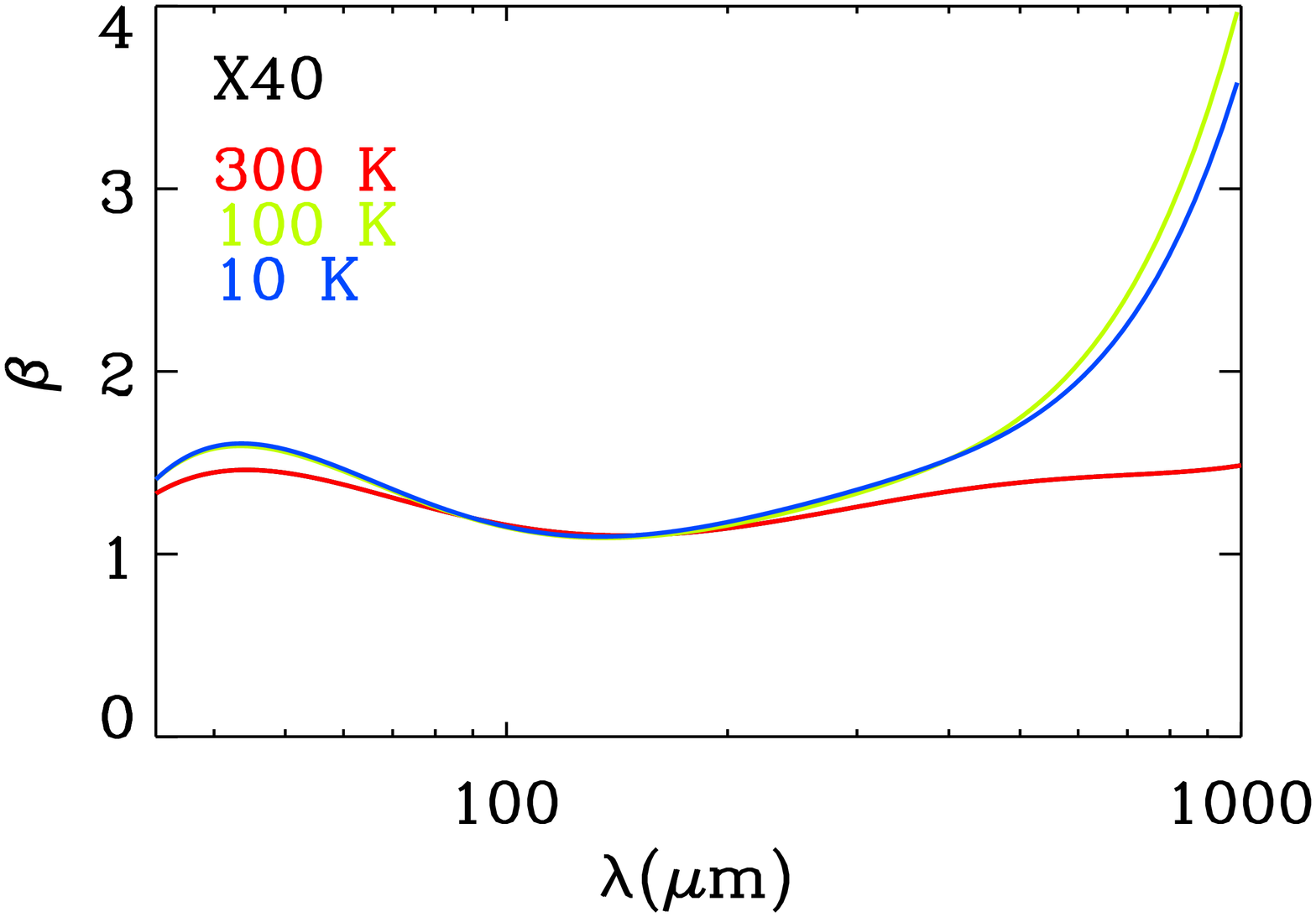}  
\includegraphics[scale=.185, angle=90]{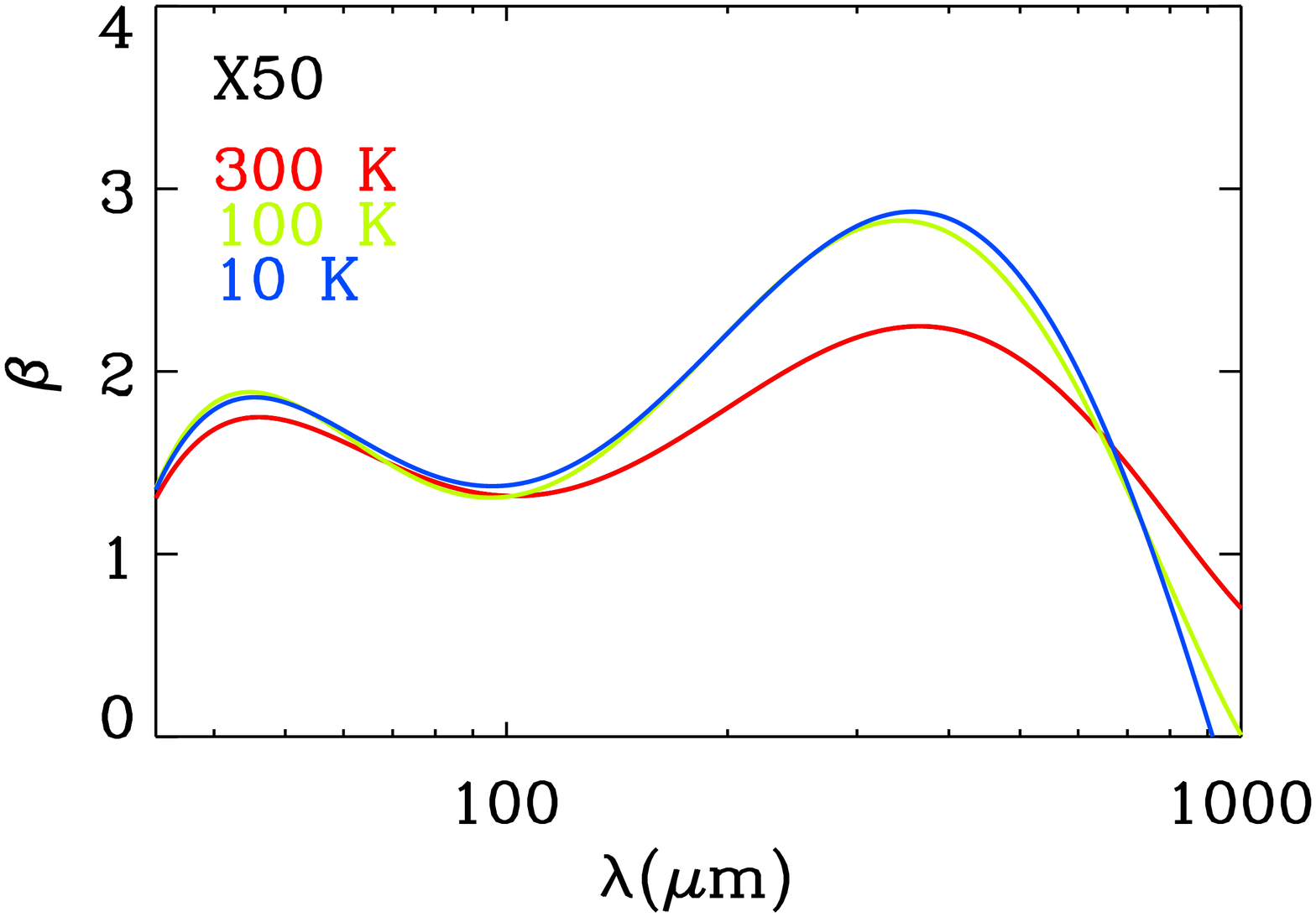}  
\includegraphics[scale=.185, angle=90]{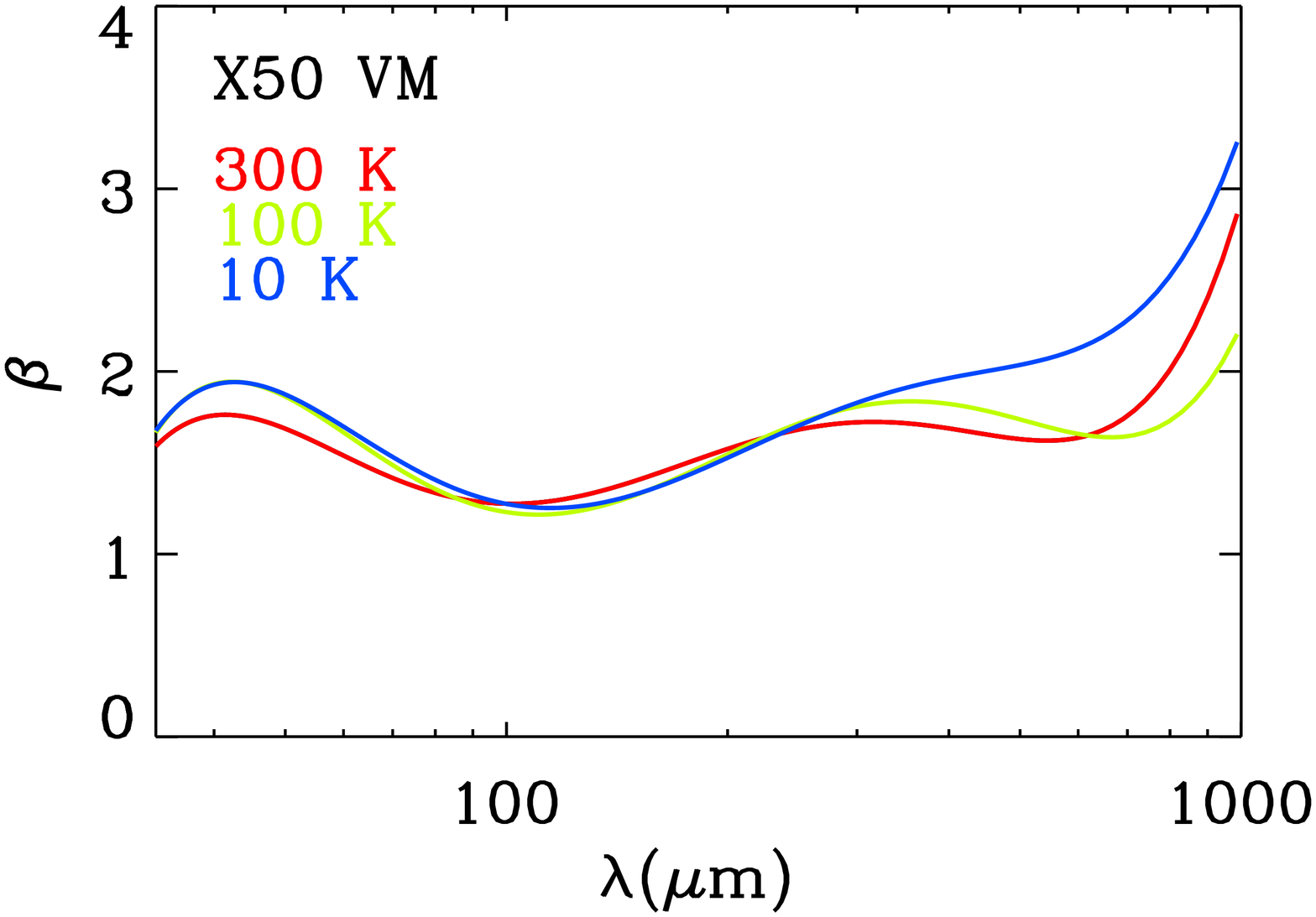}  
       \caption{Spectral index as a function of the wavelength for the studied samples: X35 {\it (upper left panel)}, X40 {\it (upper right panel)}, X50A  {\it (lower left panel)} and X50B  {\it (lower right panel)}. For each sample, the spectral index is displayed for different temperatures: 300 K (red), 100 K (green) ans 10 K (blue). The value of the spectral index was derived from a 6$^{th}$ order polynomial fit of the measured MAC. }
    \label{beta}
\end{figure}

Figure~\ref{kappa_all} presents the MAC of the studied samples in the whole spectral range measured and at 10, 100, 200 and 300 K. The MAC measured at 30 K is similar to that at 10 K. In the FIR, all the spectra show the now well-known temperature dependence characteristic of amorphous materials. {\it (i)} The MAC decreases with decreasing grain temperature. {\it (ii)} Consequently the spectral index (defined as the local slope of the spectrum) varies with the temperature: {\it (iii)} at a given wavelength, it increases when the temperature decreases. This is evident in Fig.~\ref{beta} which shows the value of $\beta$ calculated at each wavelength. This temperature effect is observed when the temperature difference between the measurements is high enough, that is, for the spectra measured at 300, 200, 100 and 10K. However we observe no temperature variation of the MAC between 10 K and 30K in the studied spectral range. The MAC variations with the temperature are more pronounced at long wavelengths. They become detectable above different wavelengths according to the samples: from $\sim$ 300 $\mu$m for the X35 samples up to 500 $\mu$m for the X40 and X50B samples (Fig.~\ref{kappa_all}). Below these wavelengths, the MAC is more or less invariable from room temperature to 10 K. In addition, for each sample, the wavelength at which these variations are detectable, that is, the wavelength at which the temperature activated absorption appears, depends on the grain temperature: it decreases with increasing temperature of the sample. In addition, to the T-$\beta$ anticorrelation, Fig.~\ref{beta} also illustrates the wavelength dependence of the MAC of the samples. It is clear from Fig.~\ref{beta} that the measured MAC cannot be fitted with a power law and a single spectral index over the whole spectral range and in particular above 100 $\mu$m, in the spectral domain where the MBB models is usually adopted in astronomical studies. 

\begin{figure*}[!t]
\includegraphics[scale=.4]{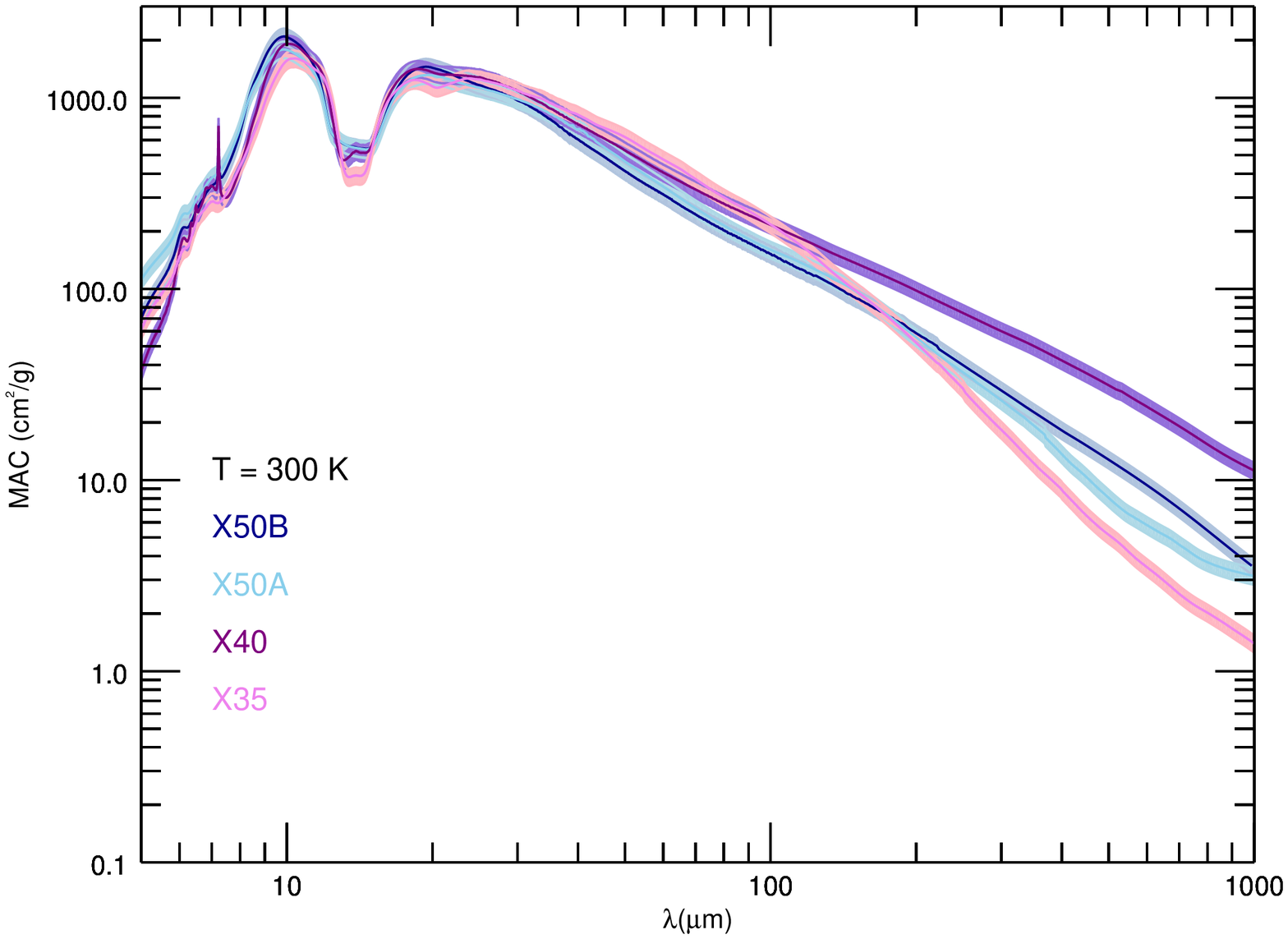}  
\includegraphics[scale=.4]{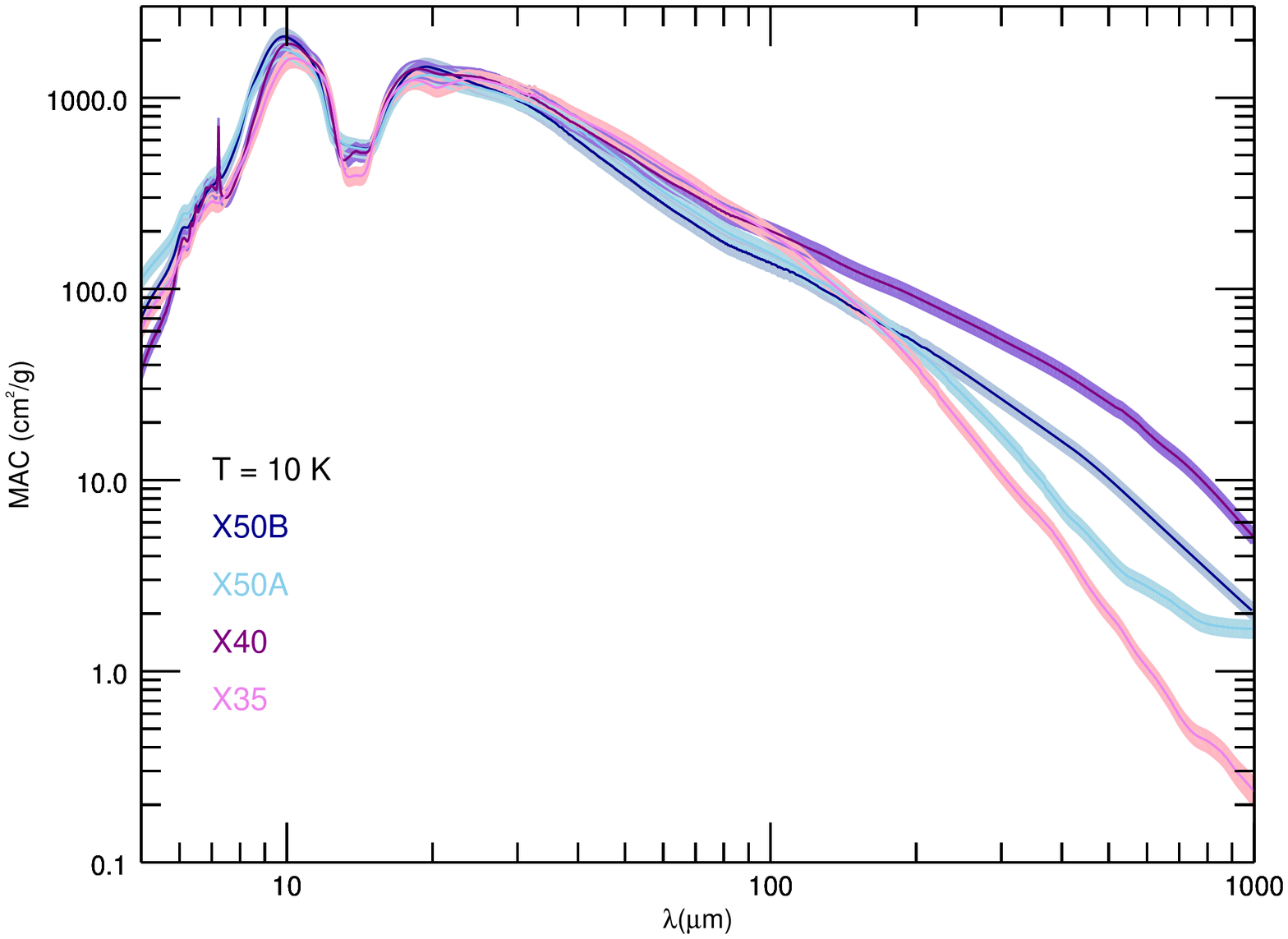}  
      \caption{Mass absorption coefficient of the samples X35, X40, X50A and X50B at room temperature  {\it(left panel)} and at 10K {\it(right panel)}. The sharp band at 7.2 $\mu$m is not associated with the samples and is due to the PE matrix.}
    \label{kappa_comp}
\end{figure*}

The MAC of the four samples are compared at 10 K and 300 K in Fig.~\ref{kappa_comp}. Up to 100 $\mu$m all MAC are similar, as can be seen in Fig.~\ref{kappa_mir}. Above 100 $\mu$m, the MAC of the four samples present different spectral shapes and the differences are more pronounced at 10 K than at 300 K. The X35 sample is the least absorbent whereas the X40 sample is the most absorbent. The two X50 samples have very similar spectra up to 300 $\mu$m at 300 K and up to 200 $\mu$m at 10 K but are different at longer wavelengths. In the FIR, as in the MIR, the spectrum of amorphous silicates depends on the structural state of the material. However it is also affected by the presence of defects in the material. Hence, these differences might come from the fact that the X50A and X50B samples, being synthesized using different methods, do not contain the same distribution of defects or the same type of defects. 

\section{Comparaison with previous experimental data}
\label{comp_lab}

As suggested, for a given composition, the structure at nanometer scale of the produced analogues depends on the methods used for the synthesis. In that sense, each sample is unique and has a structure different from the others. Glassy materials produced by quenching of high temperature liquids have a higher degree of polymerization of the silicates network than materials produced via solgel methods. Solgel samples are characterized by their water content (which can be high) and by their significant porosity due to a large amount of defects in their structure. They thus have higher specific surface and contains a lot of dangling bonds. On the contrary, quenched glasses contain almost no water, have fewer defects and are more compact than solgel samples. All these structural differences may impact the spectral properties of the material. We must use caution when relating the spectroscopic and physical properties of a given material. There are several reasons for this. During the synthesis process it is difficult to precisely control the stoichiometric homogeneity at nm-scale, the porosity and defects present in the sample. In addition these quantities are difficult to measure.

Despite the differences of the dust analogues retained in previous studies \citep{mennella1998, boudet2005, coupeaud2011} and in the present one, the spectroscopic properties and temperature dependent behavior of all these analogues remain qualitatively similar: the MAC value is correlated with the dust temperature and has a complex spectral shape differing from a simple asymptotic behavior in ${\lambda}^{-2}$. At the same time, because of these differences between the samples in terms of internal structure, composition, porosity, homogeneity, the shape of the MAC in the FIR varies. Consequently, the MAC value at a given wavelength of all samples from this study and from previous studies spans a large range of values (see Table~\ref{table_kappa} for the MAC value of the samples studied here and Table 1 from \cite{demyk2013} for previous studies at 1 mm). It must be noted that the MAC in \citet{boudet2005} has been corrected for grain shape effect whereas in \citet{coupeaud2011} and \citet{mennella1998} and in the present study this correction was not applied (see Sect~\ref{data_red}). This partly explains why the MAC value are smaller in Boudet's study, together with the effect of differences in the samples in terms of composition and structure. At 100 $\mu$m, the MAC values of all the samples from this study and from previous work are similar at 300 K and 10 K (50 - 300 cm$^2$.g$^{-1}$ at 300 K and 40 - 260 cm$^2$.g$^{-1}$ at 10 K). At 500 $\mu$m, the MAC is greater at 300 K  than at 10 K (2 - 20 cm$^2$.g$^{-1}$ at 300 K and 0.7 - 13 cm$^2$.g$^{-1}$ at 10 K). This is also the case at 1 mm where the MAC is in the range 0.1 -11  cm$^2$.g$^{-1}$ at 300 K and 0.12 - 7 cm$^2$.g$^{-1}$ at 10 K. We note that for each sample the MAC at 300K is always greater than at 10 K. Furthermore, the fact that the range of values of the MAC at 300 K and 10 K overlaps results from the dispersion of the measured spectra. This diversity of the MAC value and spectral shape that vary as a function of the materials studied can be problematic for astrophysical applications since one has to decide what analog is the most relevant and should be used in the modeling. This is considered in more detail in Sect~\ref{discuss}.   \\

\begin{table*}[!t]
\caption {Value of the MAC of the glassy samples X35, X40, X50A and X50B.} 
\label{table_kappa}
\begin{center}
\begin{tabular}{llc|cc|cc|cc|cc}
\hline 
\hline 
  & \multicolumn{10}{c}{Mass Absorption Coefficient (cm$^2$.g$^{-1}$)}	\\
\cline{2-11}  &  \multicolumn{2}{c}{100 $\mu$m}	 &  \multicolumn{2}{c}{250 $\mu$m}	 &  \multicolumn{2}{c}{500 $\mu$m}	& \multicolumn{2}{c}{850 $\mu$m} &    \multicolumn{2}{c}{1 mm}	\\ 
 	&   10K 	& 300K &   10K 	& 300K &   10K 	& 300K	&   10K & 300K	 &     10K & 300K	 \\ 
\hline 
X35	 &  194.6 & 217.2 & 	19.2 & 30.1 & 	2.0 & 5.2 &  	0.4 & 1.8 &	0.2 & 1.4   \\
X40	 &  200.9 & 216.9  & 	68.6 & 74.7 &	25.4  & 31.3  & 	7.9 & 14.0 &	 5.0	& 11.2   \\
X50A & 153.4 & 160.2  & 	27.9 & 36.5 & 	3.9 & 8.1 & 	1.7 & 3.5 & 	1.7	& 3.1	  \\
\vspace{0.1cm}
X50B & 136.8 & 152.2  & 	36.5 & 39.9 & 	10.2 & 12.8   & 	2.9 & 4.8 &	2.0 & 3.5	  \\
 < MAC1 >\tablefootmark{(1)} & 171.4	& 186.4 & 	38.1 & 45.3 &	10.4 & 14.3 & 	3.2 & 6.03 &	2.2 	& 4.8	  \\
\vspace{0.1cm}
 < MAC2>\tablefootmark{(2)} & 169.8	& 186.7 & 	25.7 & 34.1 &	4.5 & 7.8 & 	1.3 & 3.0 &	1.0 	& 2.3	  \\
MAC sphere 0.1 $\mu$m \tablefootmark{(3)} & \multicolumn{2}{c|}{33.5} & \multicolumn{2}{c|}{5.1} & \multicolumn{2}{c|}{1.2} & \multicolumn{2}{c|}{0.48} & \multicolumn{2}{c}{0.36}\\
MAC distrib spherical grains \tablefootmark{(3)} & \multicolumn{2}{c|}{36.9} & \multicolumn{2}{c|}{4.9} & \multicolumn{2}{c|}{1.1} & \multicolumn{2}{c|}{0.45} & \multicolumn{2}{c}{0.34}\\
MAC distrib prolate grains \tablefootmark{(3)} & \multicolumn{2}{c|}{50.3} & \multicolumn{2}{c|}{6.8} & \multicolumn{2}{c|}{1.5} & \multicolumn{2}{c|}{0.6} & \multicolumn{2}{c}{0.47}\\
\vspace{0.1cm}
 MAC CDE\tablefootmark{(3)} & \multicolumn{2}{c|}{74.0} & \multicolumn{2}{c|}{11.5} & \multicolumn{2}{c|}{2.6} & \multicolumn{2}{c|}{1.05} & \multicolumn{2}{c}{0.81}\\
\hline
\hline 
\end{tabular} 
\tablefoot{ \\
\tablefoottext{1}{<MAC1> represents the MAC averaged over the four samples X35, X40, X50A and X50B, see Sect.~\ref{dust_model}.\\} 
\tablefoottext{2}{<MAC2> represents the MAC averaged over the sample X35 and the mean of the X50A and X50B samples, see Sect.~\ref{dust_model}.\\} 
\tablefoottext{3}{These MAC are calculated using the optical constants of the "astrosilicates" from \citet{li2001}, see Sect.~\ref{dust_model}.\\} 
}
\end{center}
\end{table*}

The comparison of the four samples at 300 K and 10 K (Fig.~\ref{kappa_comp}) shows that, above 100 $\mu$m and at any temperature, the MAC are very different from one sample to the other. In their study of solgel Mg-rich silicates dust analogues, \citet{coupeaud2011} propose that the spectral shape of the MAC depends primarily on the cation/Si ratio and that the temperature variations of the MAC may be related to the connectivity of the SiO$^{4-}_4$ tetrahedra. From the values of the ratio of the MAC at high temperature (300 K, 200 K, 100 K) to the MAC at low temperature (10 K), we find that the amplitude of the variation of the MAC with temperature increases from the most polymerized samples (X50A and X50B) to the least polymerized sample (X35). In the wavelength range 800-1000 $\mu$m, the MAC of the X35 sample (which contains only isolated SiO$^{4-}_4$ tetrahedra and bridged Si$_2$O$_7^{6-}$ dimeric structures, see Sect.~\ref{studied_samples}), varies by a factor three to four from 10 K to 200 K as compared to a factor 1.3 to 1.6 for the X40 sample (which contains isolated tetrahedra, dimers and chains), and from 1 to 1. 4 for the two X50 A and X50B samples (which contains mainly chains and three-dimensional structures). These results are consistent with the interpretation that the temperature-activated processes responsible for the MAC variations are linked to the structural state of the material \citep{meny2007}. The connectivity or polymerization of the SiO$^{4-}_4$ tetrahedra, estimated by the population of Q$^{\mathrm{i}}$,  seems to be the most appropriate parameter for describing the material structural state. \\


\section{Discussion}
\label{discuss}

\subsection {Comparaison with cosmic dust models}
\label{dust_model}

\begin{figure}[!t]
\includegraphics[scale=.38, angle=90]{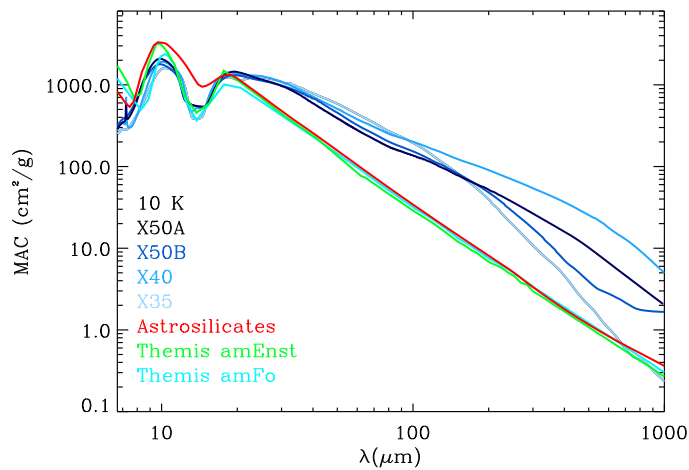}  
       \caption{Comparison of the measured MAC of the samples X50A, X50B, X40, X35 at 10K with the MAC calculated from cosmic dust models using Mie Theory for spherical grain of 0.1 $\mu$m size: "astrosilicates"  from \citep{li2001} (red curve) and the two silicates from the "Themis" dust model \citep{jones2013}: amorphous enstatite (green curve) and amorphous forsterite (cyan curve). }
    \label{kappa_comp_model}
\end{figure}

In order to compare the measured MAC with cosmic dust models we have calculated the MAC of their silicate component either using Mie theory for spherical grains or using the DDA code DDSCAT 7.3 developed by \citet{draine2013} for prolate grains (with an axis ratio of two). We calculated the MAC for different cases: for a spherical grain of 0.1 $\mu$m diameter, typical of interstellar silicate size; for two populations of grains characterized by a log-normal distribution of size with a mean size of 1 $\mu$m (to represent the grain size distribution of the studied samples) and for two different grain shapes, spherical grains and prolate grains with an axis ratio of 2; and for a continuous distribution of ellipsoids (CDE model which assumes that all ellipsoidal shapes are equiprobable, \citet{bohren1998}). For comparison of the MAC intensity, it is best to compare with the MAC modeled for a distribution of ellipsoids as it is the closest to the experimental grains distribution within the pellets. 

Figure~\ref{kappa_comp_model} shows the comparison between the MAC of the measured samples and the MAC of silicates from astronomical dust models: the "astrosilicates" from \citet{li2001} and the silicates from the "Themis" model \citep{jones2013}. In the MIR range (2.8 - 17.5 $\mu$m), the "Themis" model is based on the laboratory work from \citet{scottduley1996}, that is, on amorphous enstatite (MgSiO$_3$) and forsterite (Mg$_2$SiO$_4$) produced by laser ablation. The optical constants of the "astrosilicates"  are constructed from observations of dust emission in the Trapezium region for the 9.7 $\mu$m band and from dust emission from circumstellar dust shells around O-rich stars as well as from dust absorption toward the Galactic Center  for the 18 $\mu$m band \citep{draine1984}. In the MIR domain, up to $\sim$ 20 $\mu$m the laboratory spectra and the "Themis" modeled spectra agree reasonably well in terms of band shape and MAC value, even though differences are expected since the samples on which the "Themis" model is based are different from the four glassy samples. The comparison with the "astrosilicates" shows that the stretching mode at $\sim$ 9.7 $\mu$m of the "astrosilicates" is broader and more intense than the one of the studied analogues, independently of assume grain size and shape distribution adopted for the model. Because the "astrosilicates" are constructed from astrophysical observations, these differences cannot be discussed in terms on material properties. They may be explained by the fact that additional silicate components are needed to reproduce the line of sight from which the "astrosilicates" are constructed. It is known that the observed MIR silicate bands vary with the environments and that no unique silicates, from laboratory data or derived from observations, can reproduce the MIR silicate bands in all astrophysical environments \citep[see][and reference therein]{fogerty2016}. It is thus important to emphasize that the observed spectral differences in the MIR do not question the relevance of the studied analogues. 



At longer wavelengths and in the temperature range of astrophysical interest 10 K - 30 K (where T-independent processes fully dominate the absorption) the laboratory spectra and the modeled ones are very different. The measured MAC for the X40, X50A and X50B samples are much higher than the modeled MAC in the whole spectral range. This is also true for the X35 samples for $\lambda \le$ 700 $\mu$m but for $\lambda \ge$ 700 $\mu$m the measured MAC is similar to the modeled one. The increase of the measured MAC with respect to the MAC modeled for spherical grains can be calculated at some specific wavelengths from Table~\ref{table_kappa} (as well as for other shapes and size distributions). The X40 sample, the most absorbent, is 4 times more absorbent at 100 $\mu$m than the models and about 10 to 17 times at longer wavelengths. The two X50 samples are 3 to 7 times more absorbent than the models over the entire spectral range. For the X35 samples the MAC is 1 to 3 times more absorbent in the 100 - 600 $\mu$m range and becomes as absorbent or even less absorbent at longer wavelengths. At wavelengths greater that 20 $\mu$m, the "astrosilicates" and the "Themis" silicates are not based on experimental measurements or on observations but on the extrapolation of the optical constants derived in the MIR. The differences between the measured MAC and the MAC from cosmic dust thus do not reflect differences in terms of material but rather in terms of theory of the absorption processes (see Sect.~\ref{abs_process}).  \\

As highlighted in Sect.~\ref{comp_lab} and as seen in Fig.~\ref{kappa_comp_model}, the diversity of the MAC from one sample to another raises the question of the relevant dust analogues to be used in astronomical models. To investigate this question it is interesting to consider what we can learn from the many (about one thousand) presolar silicate grains that have been identified \citep[and references therein]{nguyen2016}. The detailed study, using Transmission Electron Microscopy (TEM), of these presolar silicates grains shows that they have a wide range of microstructure, shape and composition. They are mostly amorphous and often non-stoichiometric with sometimes internal nm-scale compositional variations \citep{nguyen2016}. \cite{bose2012} found in the CO3 chondrite ALHA 77307 many presolar grains having intermediate stoichiometry between olivine-like and pyroxene-like material (see their Fig.6), similar to the X40 sample. These presolar grains come from AGB circumstellar envelops, supernovae, novae: they are the interstellar grains that emit in the FIR/submm domain. Ideally, this diversity of composition should be taken into account in interstellar dust models and the silicate component of these models should be built from various dust analogues, including non-stoichiometric analogues, for example by averaging their optical constants or MAC.  

\begin{figure}[!t]
\includegraphics[scale=.42]{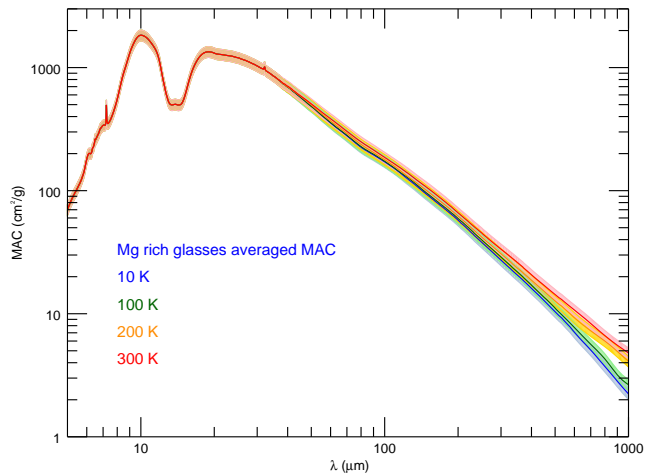}  
       \caption{Mass absorption coefficient averaged on the four samples X35, X40, X50A and X50B for a grain temperature of 300 K (red), 200 K (orange), 100 K (green) and at 10 K (blue). The shaded area represents the uncertainty on the experimental data (see text). }
    \label{kappa_mean}
\end{figure}

\begin{figure}[!t]
\includegraphics[scale=.3, angle=90]{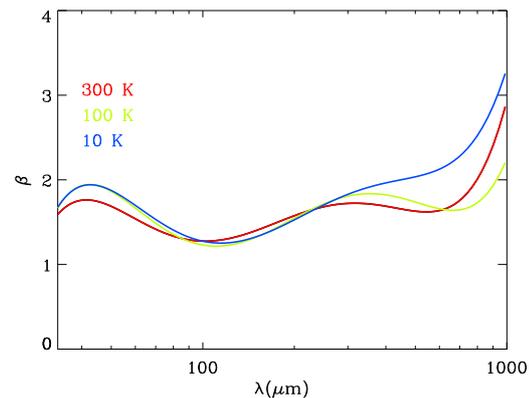}  
       \caption{Spectral index as a function of the wavelength for the MAC average over the four  studied samples, <MAC1>. The value of the spectral index was derived from a 6$^{th}$ order polynomial fit of the measured MAC. It is displayed for different temperatures: 300 K (red), 100 K (green) ans 10 K (blue).}
    \label{beta_mean}
\end{figure}

We have calculated the average MAC, <MAC1>, from the four samples X35, X40 and X50A and X50B by simply adding the four spectra and dividing the sum by four (Fig.~\ref{kappa_mean}). Such an averaged MAC has a Mg/Si ratio of 1.18, very close to the one given by the cosmic elemental abundance of magnesium and silicon which is in the range 1 - 1.15 depending on the adopted reference \citep{jenkins2009,przybilla2008}. The resulting averaged MAC presents the same characteristics as each individual sample: variation of the MAC spectral shape and intensity with the temperature. However, as expected, the amplitude of these variations are lower than those of each sample taken individually. In addition the spectral shape of  <MAC1> is smoothed by the average and the $\beta(\lambda)$ value varies on a narrower range than for each sample (Fig.~\ref{beta_mean}). At 300 K, the <MAC1> curve is well approximated by a single power law ${\lambda}^{-\beta}$ with $\beta$ = 1.6 for $\lambda \ge$ 100 $\mu$m whereas at 10K, two power laws are needed with  $\beta$ = 1.8 for 150 $\mu$m $ \le \lambda \le$ 500 $\mu$m and $\beta$ = 2.2 for $\lambda \ge$ 500 $\mu$m. If we calculate the average in terms of composition - taking the average of the X50A and X50B sample for the MgSiO$_3$ pole - the results are close to <MAC1> because the sample X40, much more absorbent than the others, dominates the average MAC. On the contrary, if we do not consider the X40 sample and averages the samples X35 and the mean of the two X50A and X50B (Mg/Si = 1.12, <MAC2>), the results are different. In this case at least four power laws are needed to reproduced the averaged MAC ($\beta$ = 1.8, 2.4, 2.3, 1.6 in the spectral ranges 100 - 180 $\mu$m, 180 - 500 $\mu$m, 500 - 800 $\mu$m and 800 - 1000 $\mu$m, respectively). 

\begin{figure}[!t]
\includegraphics[scale=.43]{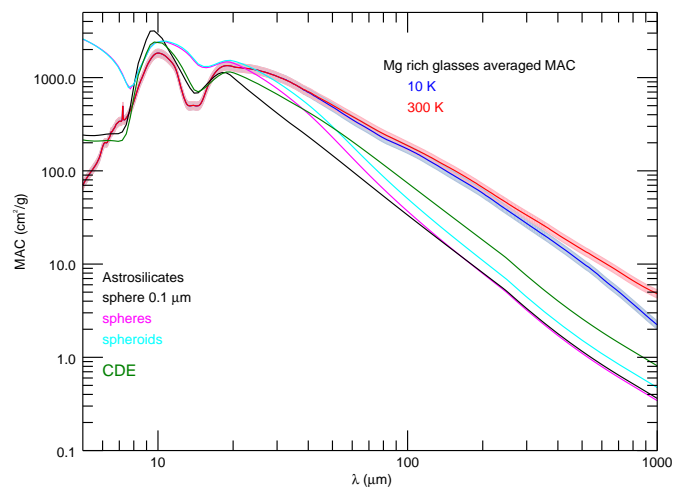}  
       \caption{Comparison of averaged experimental mass absorption coefficients with the MAC calculated from cosmic dust model. The solid line curves with the uncertainty represent the MAC averaged on the four samples X35, X40, X50A and X50B: (red) MAC at 300 K, (blue) MAC at 10K. The MAC of the "astrosilicates" from \citep{li2001} are calculated using Mie Theory for a 0.1 $\mu$m size grain (black), for a log-normal grain size distribution with a mean diameter of 1 $\mu$m for spherical grains (magenta) and for prolate grains (cyan) and for a CDE model (green). See text for more details.}
    \label{kappa_mean_comp_model}
\end{figure}

Fig.~\ref{kappa_mean_comp_model} shows that the averaged MAC, <MAC1>, is much more absorbing than that MAC from cosmic dust models. The comparison of the MAC of cosmic dust models with <MAC1> shows that at $\lambda$ = 100 $\mu$m the measured MAC is 3.4 times larger than the dust model calculated for a size distribution centered at 1 $\mu$m of prolate grains (Table~\ref{table_kappa}). At $\lambda$ = 1mm <MAC1> is 4.6 times larger than the cosmic silicates dust models for grain temperature of 10 K. The comparison with the CDE model leads to a factor of 2.3 and 2.7 respectively at 100 $\mu$m and 1 mm for <MAC1>  measured at 10 K. If one compares with <MAC2>, then, at 10K, <MAC2> is higher than the modeled one by a factor of 3.5 at 100 $\mu$m and by a factor of 2.5 at 1 mm. \\


\subsection{Absorption processes}
\label{abs_process}

The difference between the MACs calculated from cosmic dust models and from experimental data has important astrophysical implications in terms of mass and temperature determination which needs to be understood. It could arise from experimental artefacts (such as the effect of size distribution, the effect of coagulation within the pellets) or it may be due to absorption processes intrinsic of the material. In the following, we investigate these possibilities.

The measured MACs correspond to a population of grains with a certain shape (not spherical) and size distribution (unknown) and this has to be taken into account when comparing with the MAC calculated from dust models. The typical sizes which can be reached by manually grinding minerals in an agate mortar are around 1 $\mu$m, size that was adopted for the mean size for the size distribution. Increasing the grain size broadens the vibrational bands and changes their peak position and relative strength. The stretching and bending modes thus constrain the amount of large grains and the width of the chosen size distribution. However, despite these constraints, the solution is degenerated and several grain size distributions can give a reasonable fit to the experimental spectra in the MIR range. However, none of these size distributions can explain the discrepancy between astronomical dust models and our experimental data for $\lambda \geq$ 70 $\mu$m, although size distribution effects may contribute to the increase of the MAC value and to the broadening of the red wing of the bending O-Si-O vibrational mode in the 20-70 $\mu$m range (Fig.~\ref{kappa_mean_comp_model}). This conclusion holds for a distribution of non spherical grains that does not change much the MAC compared with spherical grains, although it slightly increases the MAC at $\lambda \geq$ 100 $\mu$m (Fig.~\ref{kappa_mean_comp_model}). 

A second effect is the aggregation of sample grains within the pellets. This phenomenon may occur when preparing the pellets but it is difficult to control during fabrication and impossible to probe afterwards. This effect is in principle taken into account in the data reduction analysis thanks to an Effective Medium Theory approach. The corresponding correction, which applies to the whole spectrum, is small (a factor in the range 0.88 - 1, see Sect.\ref{studied_samples}). Using more sophisticated modeling approaches like the DDA (Discrete Dipole Approximation) approach, it has been shown that the maximum factor of enhancement due to coagulation should not exceed two, whatever computational method used, and whatever shape, compactness and size of the aggregates are considered \citep{stognienko1995,koehler2011,mackowski2006,min2016}. Thus, although coagulation may occur within the pellet it cannot explain enhancement factors greater than three as observed nor the different enhancement factor at 1 mm relative to the one at 100 $\mu$m. This is because coagulation of grains of identical material do not change the slope of the MAC, meaning that the enhancement factor does not vary with wavelength, in the FIR/submm. \\

All this indicates that for $\lambda \ge$ 100 $\mu$m our samples are intrinsically more absorbent than the silicates used in cosmic dust models. For $\lambda \ge 20 \mu$m, both dust models use the Debye model to extrapolate the optical constants of the material. The MAC thus has an asymptotic behavior in ${\lambda}^{-2}$. However, several other absorption processes may occur in the FIR range. In the TLS model, built to explain astronomical observations at $\lambda \ge 100 \mu$m where the emission of big grains dominates the interstellar radiation field \citep{meny2007}, two absorption processes are considered, the DCD and the TLS processes. The DCD model \citep[Disorder in Charge Distribution, ][]{schlomann1964} describes the FIR absorption in a disordered charged distribution simply characterized by the correlation length $l_{\rm c}$ that represents the averaged length over which charge neutrality is broken in the disordered material. Such a model induces a dependence of the MAC in $\lambda^{-2}$ in the vicinity of  $\lambda$ >100  $\mu$m and an asymptotic behavior in the long wavelength limit in $\lambda^{-4}$. It is a temperature-independent absorption. The TLS \citep[Two Level System,][]{phillips1987} component of the model describes temperature activated absorption processes which are added to the DCD absorptions. In this model the atomic disordered network is described by an ordered distribution of asymmetric double potential wells (the TLS), each of which can be viewed as two configurations of atoms or groups of atoms. Transitions within each TLS can occur via resonnant or relaxational temperature-dependent processes \citep[see][and reference therein for more details]{meny2007}.

In the absence of experimental measurements of the MAC in the wavelength range 25 - 100 $\mu$m in previous studies, the TLS model was simply expected to slightly modify the slope of the FIR/submm MAC of the silicates calculated from astronomical dust models. In particular, the absolute value of the MAC at $\lambda = 100$ $\mu$m was expected to be the same for the amorphous silicates and for the astronomical dust models. However the present measurements show that in the wavelength range from 10 to 100 $\mu$m, the MAC is more complex than expected from simple semi-classical models (Lorentz and Debye) or with TLS models.

The larger value and the spectral shape of the experimental MAC may be related to the so-called Boson peak (BP) which appears to be a universal feature of solid state of very different amorphous materials such as oxide glasses, amorphous ices, polymers \citep{champagnon2009}. The BP has been widely observed in the FIR in Raman spectroscopy \citep{kojima2000,kalampounias2009}, inelastic neutron-scattering spectroscopy \citep{buchenau1984,nakayama2002}, terahertz time-domain \citep{kojima2005,parrott2010} or a combination of reflection and transmission technics and spectroscopy in the FIR/submm \citep{schneider2001,parrott2010}. However the exact origin and nature of the BP has been the subject of many debates over more than three decades. Several attempts, frequently self-contradictory, have appeared so far for the explanation of the BP in non-crystalline solids. These can be grouped into two main categories. On one side, the BP is considered to arise from localized or quasi-localized modes; the existence of weak interatomic forces into the glass create the so-called soft potential \citep{klinger1988,gurevich2003}. From an other point of view, the BP is a manifestation that the disorder impacts on the vibrational density of states (VDoS) of the corresponding crystalline solid. Depending on the composition of the glass, the BP has been observed in a broad optical wavelength range, extending roughly from $\lambda$ = 100 to 1000 $\mu$m. 
 
Very few physical and analytical models exist that describe BP observations. In the analytical model developed by \citet{gurevich2003}, the BP originates from quasi-local vibrations (QLVs) characterized by large vibrational amplitudes of some groups of atoms, which is a typical feature of disordered systems and very similar to the well known TLS. These QLVs can be modeled as local low frequency harmonic oscillators coupled with the acoustic vibrations. In turn, this leads to dipole-dipole interactions between the different harmonic oscillators. The important results of this model are that $g_{BP} (\omega)$, the density of states of the BP, depends on a single parameter $\omega_{BP}$  characterizing the position of the BP and that the $g_{BP}(\omega)$ excess over the $g_{Debye}(\omega)$ (the density of states of the Debye model) varies following $g_{BP}(\omega) \propto\omega^4$ at long wavelengths and $g_{BP}(\omega) \propto\omega$ at short wavelengths on both sides of the BP maximum. This model thus explains the occurence of spectral index values below the expected asymptotic value $\beta$ = 2 in the far-infrared wings of the MIR bands ($\sim$ 30 - 100 $\mu$m). The fit of the BP requires an other parameter: the amplitude of the corresponding excess of vibrations density of states over the underlying density of states, whatever Debye density of states or DCD density of states is chosen. We fit the experimental MAC curves with underlying Debye density of states with constant optical coupling coefficient, and the best fit of the measured MAC are presented in the Appendix~\ref{BP}. The fits are quite good over at least an order of magnitude in wavelength, from $\lambda$ = 30 $\mu$m for all samples, to $\lambda$ = 400 $\mu$m (sample X35),  900 $\mu$m (X40), 600 $\mu$m (X50A), 700 $\mu$m (X50B). 

It is important to keep in mind that these results are preliminary. A more detailed work, taking into account the absorption process that occur in the FIR range, will be published elsewhere. However analyzing the MAC of our samples using physical models, instead of pure empirical laws such as a power law with different spectral indexes defined over different FIR wavelength ranges appears to be important for several reasons. If a physical model can fit the data over an extended wavelength range with very few free parameters specific to the material or type of disorder, it can help our understanding and suggests new studies for solid state and theoretical physicists. In addition, such an understanding of the experimental data may provide new dust parameters to fit the observations. Most importantly this understanding may allow us to predict, within the framework of the physical model, how the astrophysical spectra evolve in more complex situations (mixtures of dust with various compositions, distribution of temperatures, amongst others). An observationnal parameter based on an empirical law does not allow such prediction. 

\subsection{Implications for astrophysical models and data interpretation}

As in previous experimental studies \citep{mennella1998,boudet2005,coupeaud2011}, the temperature dependence of the MAC of the four samples is observed for temperatures greater than 30 K. It thus cannot explain the $\beta$-T anti-correlation observed in astrophysical environments in the 10-30 K range, which, as indicated in Sect.\ref{intro}, may have several origins and reflects for example changes in the grain properties \citep{meny2007,koehler2012,paradis2011,jones2012a,jones2012b,jones2013}. The $\beta$ - T anti-correlation represents one aspect of the temperature dependence of the MAC. The second one is the decrease of the MAC value with the temperature. This has to be considered in models based on MIR-to-FIR experimental MAC since using MAC measured at room-temperature instead of 10 K induces errors greater than a factor of 2 on the MAC (see Table~\ref{table_kappa}) and thus on the calculated dust mass. This is not only true for silicate grains but also for carbonaceous grains \citep{mennella1998}. However as discussed below, using cosmic dust models extrapolated in the FIR/submm range from room temperature measurements in the MIR induces even larger error (see below). \\

The "astrosilicates" model has been modified in the FIR to reproduce the FIRAS observations showing a flattening of the dust emission \citep{draine2007} also known as the FIR excess and also observed by Herschel and Planck \citep{paradis2009,juvela2015}. The inflexion in the MAC curve is visible in Fig.~\ref{kappa_mean_comp_model} above 800 $\mu$m. The "astrosilicates" model thus considers that the FIR excess is only due to the silicate component. Such flattening of the MAC at long wavelengths is indeed observed in some samples in \citet{boudet2005} and \citet{coupeaud2011} and in this study for the X50 sample but for other samples the MAC is found to steepen at long wavelengths and this also the case for the mean MAC, <MAC1>. The FIR excess has been widely studied in our Galaxy and in external galaxies and its origin is still debated \citep{hermelo2016}. Experiments and models show that amorphous carbon matter has a flatter MAC than silicate dust \citep{mennella1998,jones2012b,jones2013} and could thus significantly participate to the excess if the grain properties in regions where the excess is observed change.  \\

The comparison with previous experimental data shows that the MAC from all studies are much higher than the MAC calculated from cosmic dust models. As discussed in Sect.~\ref{abs_process} this probably comes from absorption processes that occur in addition to the Debye model. This observation cannot be due to grain size effects and only partly to grain coagulation within the pellets. At the most, the effect of coagulation should decrease the experimental MAC by a factor 2 and consequently the values presented in Table~\ref{table_kappa} could be up to two times too high. However, even adopting this pessimistic value of 2, the MAC of the four samples are still higher than the MAC calculated from dust models: by a factor of 1.3 to 3.4 for the X50B sample depending on the wavelength, by a factor 1.3 to 2 for the X50A sample, by a factor 2 to 8 for the X40 sample, by a factor 1.4 to 2 for the X35 sample for 100 $\mu$m $\ge \lambda \le$ 500 $\mu$m. For $ \lambda$ $\ge$ 500 $\mu$m the MAC of the X35 sample is lower than the modeled one by a factor of 0.3 to 0.2. This is also true for the MAC averaged over the four samples: <MAC1> is still higher than the modeled one by a factor of 1.7 to 3.5 depending on the wavelength and for the MAC averaged over the the X35 and X50 composition: <MAC2> is still higher than the modeled one by a factor of 1.25 to 2.2 depending on the wavelength. We want to emphasize that these values are lower limits since we adopted the least favorable case for the coagulation in the pellets. \\

This has important implications for dust emission models and for the determination of dust mass and temperature. A modification of the silicate grains MAC is equivalent to a modification of their emissivity. The direct consequence for the mass determination is that masses determined from cosmic dust models are significantly overestimated compared to masses determined with experimentally measured emissivity. Considering only silicate grains in cosmic dust models, the dust mass estimated from silicate grains twice more absorbent than the "astrosilicates" or than the silicate from the "Themis" model will be twice as low. The error on the mass estimations depends on the experimental data adopted but also on the relative contribution of the silicate grains to the FIR emission, in other words to the amount of silicates grains versus large carbonaceous grains. Therefore, it appears critical to measure the MAC of carbonaceous dust analogues to put constraints on all the dust components that are thought to contribute to the FIR emission. Another consequence of having more absorbent silicate grains in cosmic dust models is that the mass of the silicate grain population needed to reproduce the observation is smaller and the constraints on the heavy element abundances of cosmic dust models is more flexible.  

To be able to incorporate these data into cosmic dust models, one needs to derive the optical constants of each sample from the MAC curves. This work is under progress. The optical constants will allow us to calculate the emissivity and scattering efficiency of these silicate dust analogues for any distribution of size and shape relevant for studies of the ISM (dense or diffuse), of circumstellar shells around evolved stars and of protoplanetary disks. It will also allow us to calculate the polarisation efficiency of this grain populations. It is important to investigate the inability of current dust models to reproduce simultaneously the polarization observations from {\it Planck} in the FIR/submm and in the visible \citep{planck-XXI-2015}. Interestingly, in this study the authors indicate that to correctly reproduce the observations, the aligned grains responsible for the polarization, that is, the silicate big grains, need to be more emissive, which is the case for the silicate analogues studied here.


\section{Conclusions}

We measured the mass absorption coefficient in the 5 $\mu$m - 1mm domain of the magnesium rich amorphous silicate dust analogues at temperature in the range of 10 K to 300 K. The samples are glassy (sub)micronic silicate grains, of compositions (1-$x$)MgO - $x$SiO$_2$ with x = 0.35 (close to Mg$_2$SiO$_4$), 0.50 (MgSiO$_3$) and 0.40 (Mg$_{1.5}$SiO$_{3.5}$). Our findings may be summarized as follow: 

\begin{itemize}

\item  We observe, for all sample compositions, that the MAC varies with the grain temperature in the FIR/submm, above a wavelength that depends on the sample ($\sim$ 200 - 500 $\mu$m) and on the sample temperature. At temperature greater than 30 K, additional absorption processes are thermally activated, resulting in an increase of the MAC with the grain temperature. These thermally activated absorption processes are characteristic of the amorphous nature and of the presence of defects within the sample structure. We do not observe temperature variations of the MAC in the 10 - 30 K range. 

\item For $\lambda \ge 100 \mu$m, the spectral shape of the MAC varies from one sample to another and cannot be approximated by a single power law such as ${\lambda}^{-\beta}$. The value of the spectral index $\beta$ varies with wavelength from one sample to another from one up to three. This is due to differences in the structure and in the type and amount of defects in the samples.

\item Above 100 $\mu$m the value of the MAC of all the samples and of the average MAC of the samples is much higher than the value of the MAC calculated from astronomical dust models. While the presence of large grains cannot account for this MAC enhancement, grain coagulation within the pellet could contribute at most by a factor two in which case the experimental MAC are still higher than the modeled MAC by a factor of two or more depending on the samples and on the wavelength. The observed enhancement of the MAC thus appears to be an intrinsic property of the grain material and in particular to the type and amount of defects within the material structure such as porosity. It is compatible with the detection of the boson peak in the spectra which induces an extra absorption with respect to the Lorentz and Debye models in ${\lambda}^{-2}$ adopted in the cosmic dust models.

\item The observed enhancement of the MAC has important consequences on astrophysical modeling of dust (dust mass and temperature). The dust mass estimated with dust models are overestimated with respect to the estimation that would be made using experimental values presented in this study.

\end{itemize}


\begin{acknowledgements}
This work was supported by the French \emph{Agence National pour la recherche} project ANR-CIMMES and by the program Physique et Chimie du Milieu Interstellaire (PCMI) funded by the Conseil National de la Recherche Scientifique (CNRS) and the Centre National d'études Spatiales (CNES) of France. X. Lu aknowledges financial support from the French Region Midi-Pyrénées. We thank A. Retsinas for technical support and N. Ysard,  A. Jones and H. Mutschke for fruitful discussions.
\end{acknowledgements}

\bibliographystyle{aa} 
\bibliography{bib} 

\appendix
\section{Physical modeling of the MAC}
\label{BP}

This appendix presents the figures showing the modeling of the MAC curve of each sample using the Debye and the BP models as explained in Sect.~\ref{abs_process}. We used the BP model developed by \citet{gurevich2003}. The spectral region below 30 $\mu$m is not considered in the fit. Each figure shows the measured MAC, the best model, the Debye component of the model and the BP component of the model. 

\begin{figure}[h]
\includegraphics[scale=.37 ]{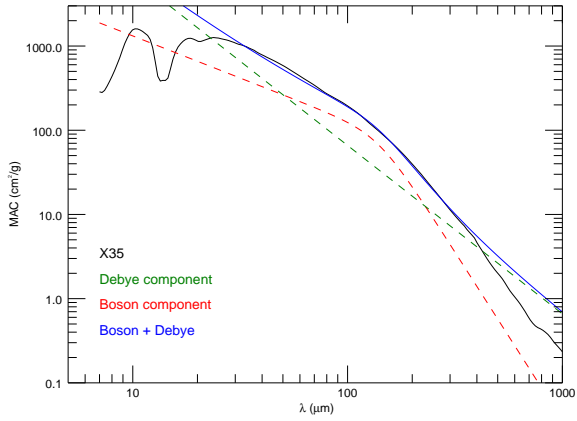}  
       \caption{Modeling of the MAC of sample X35. The figure shows measured MAC (black line), the best model (blue line), the Deby component of the model (green dotted line) and the bP component of hte model (red dotted line).}
    \label{BP_fit_x35}
\end{figure}

\begin{figure}[h]
\includegraphics[scale=.37]{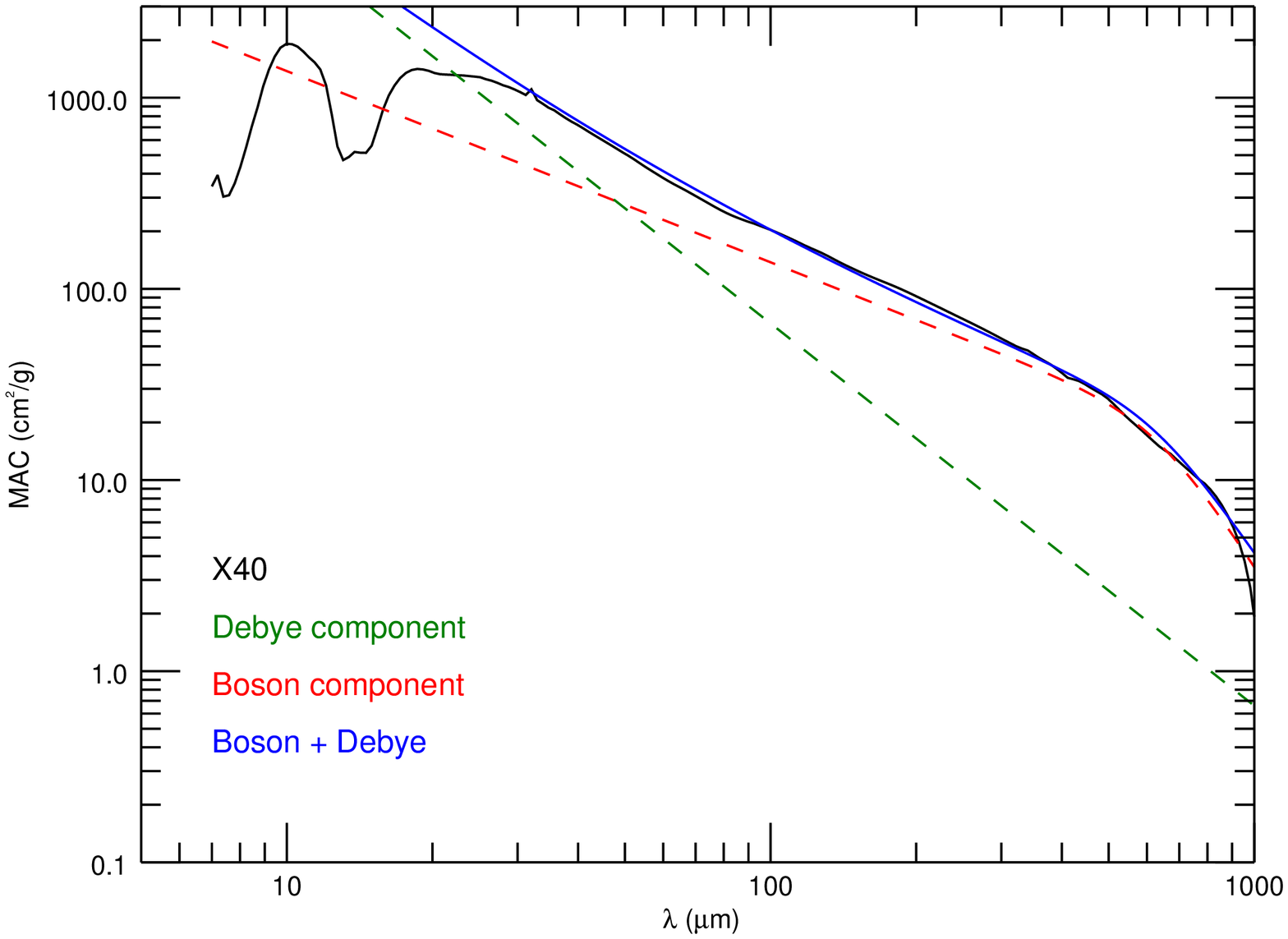}  \\
       \caption{Modeling of the MAC of sample X40. The figure shows measured MAC (black line), the best model (blue line), the Deby component of the model (green dotted line) and the bP component of hte model (red dotted line).}
    \label{BP_fit_x40}
\end{figure}

\begin{figure}[h]
\includegraphics[scale=.37]{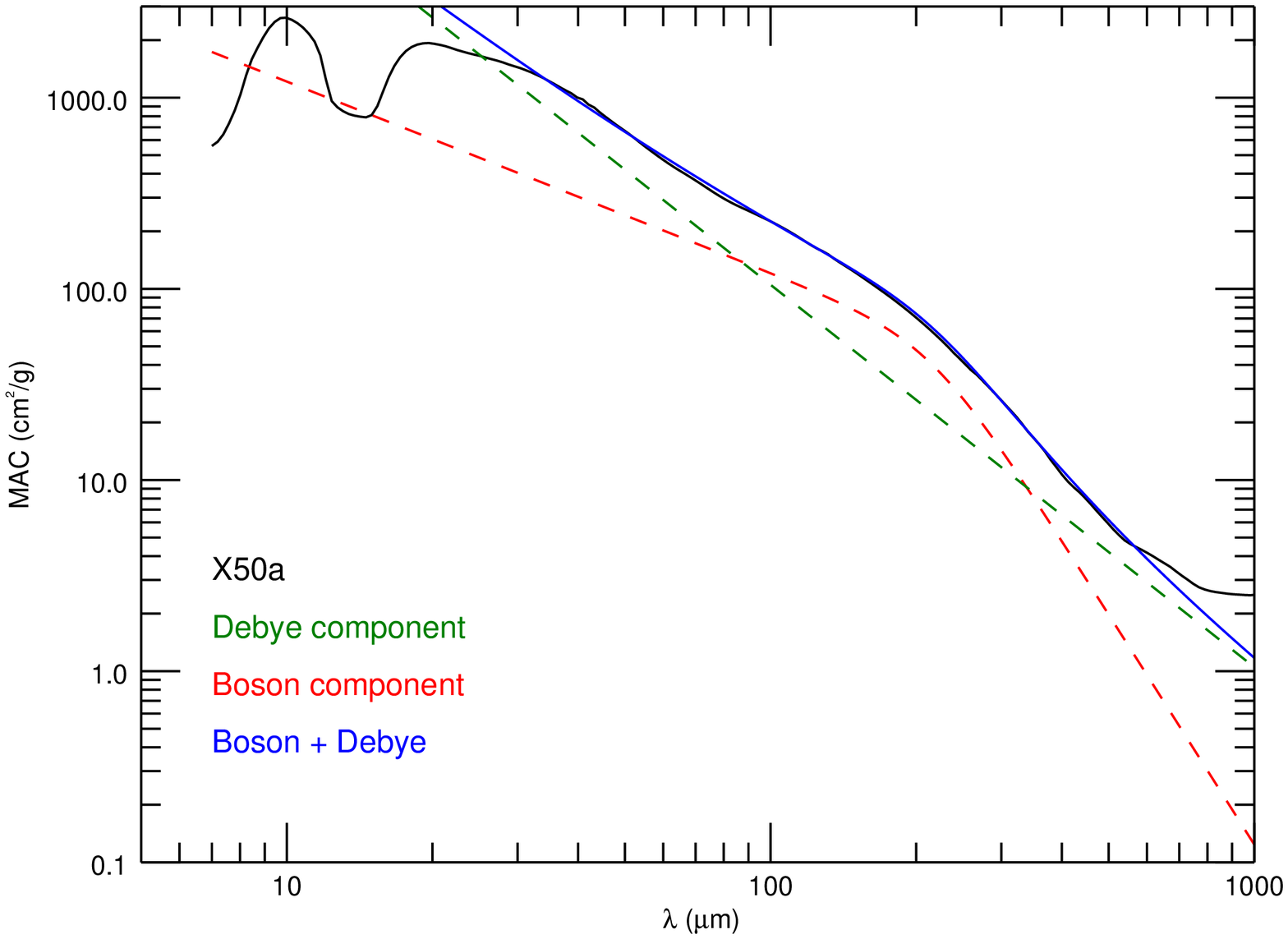}  
       \caption{Modeling of the MAC of sample X50A. The figure shows measured MAC (black line), the best model (blue line), the Deby component of the model (green dotted line) and the bP component of hte model (red dotted line).}
    \label{BP_fit_x50}
\end{figure}

\begin{figure}[h]
\includegraphics[scale=.37]{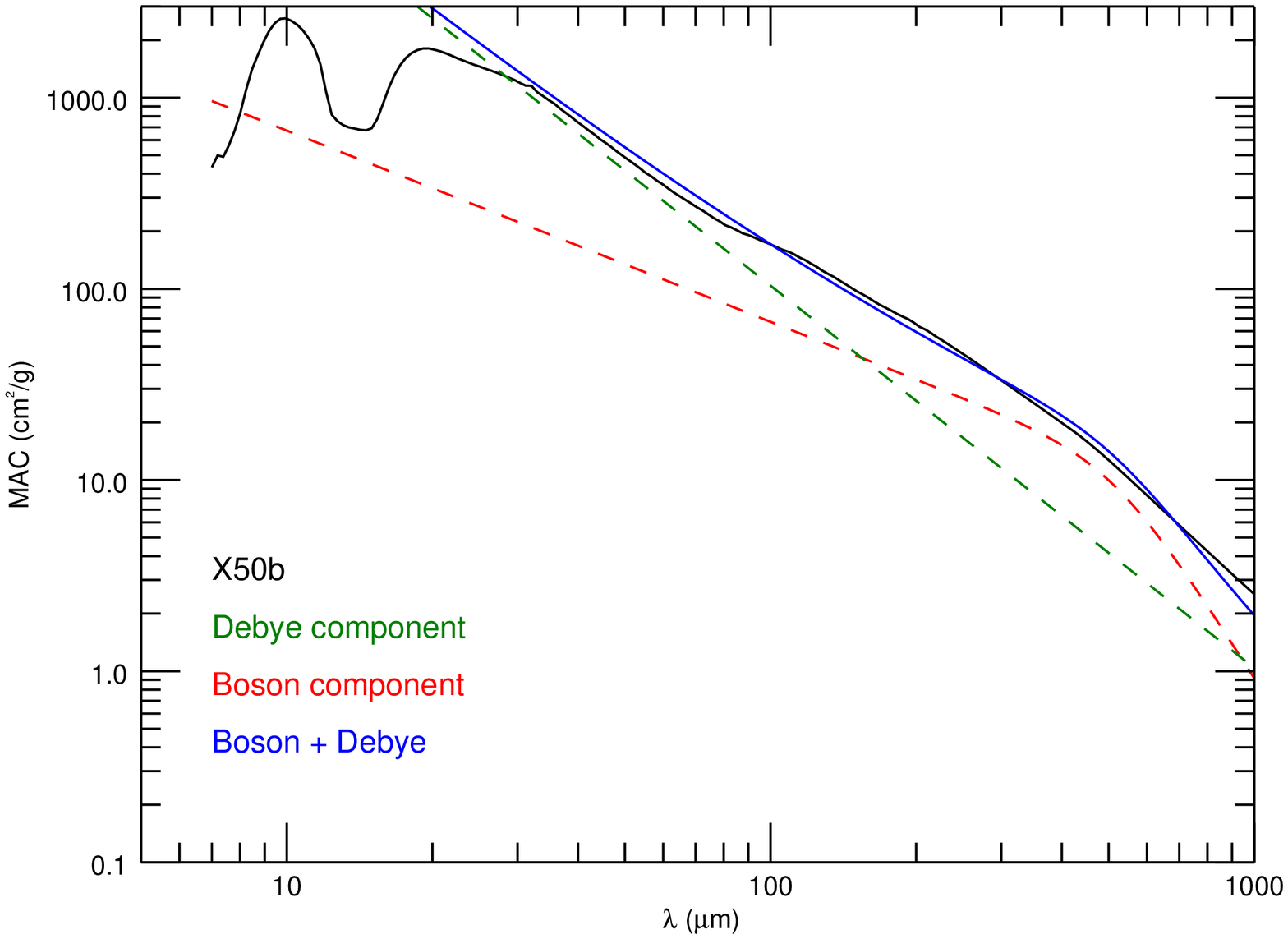}  
       \caption{Modeling of the MAC of sample X50B. The figure shows measured MAC (black line), the best model (blue line), the Deby component of the model (green dotted line) and the bP component of hte model (red dotted line).}
    \label{BP_fit_x50}
\end{figure}

\end{document}